\pdfoutput=1
\def\preprint{}

\def\PreprintAuthors{%
  Xinyu Wang$^{1}$\thanks{Equal contribution.} \quad
  Ziyu Zhao$^{2,1}$\footnotemark[1] \quad
  Yixuan He$^{2}$ \quad
  Xiao-Wen Chang$^{1}$ \quad
  Alex Smola$^{3}$\thanks{Corresponding author.} \\
  $^{1}$McGill University \quad $^{2}$Arizona State University \quad $^{3}$BosonLabs}

\documentclass{article}
\usepackage{iclr2027_conference,times}

\usepackage{amsmath,amsfonts,bm}

\def\eqref#1{equation~\ref{#1}}

\def\1{\bm{1}}

\DeclareMathAlphabet{\mathsfit}{\encodingdefault}{\sfdefault}{m}{sl}
\SetMathAlphabet{\mathsfit}{bold}{\encodingdefault}{\sfdefault}{bx}{n}

\usepackage[hypertexnames=false]{hyperref}
\hypersetup{hidelinks}
\usepackage{url}
\usepackage{amsmath,amssymb}
\usepackage{booktabs}
\usepackage{array}
\usepackage{amsthm}
\newtheorem{proposition}{Proposition}

\newtheorem{corollary}[proposition]{Corollary}
\theoremstyle{remark}

\theoremstyle{plain}
\usepackage{graphicx}
\usepackage{float}
\usepackage{placeins}
\graphicspath{{figures/}}

\newcounter{noteYHctr} 

\title{When Is Deletion Ordering Tractable? From Update Dynamics to Permutation Structure}

\ifdefined\preprint
\author{\PreprintAuthors}
\iclrfinalcopy
\else
\author{Anonymous authors\\Paper under double-blind review}
\fi

\begin{document}
\maketitle
\ifdefined\preprint\lhead{Preprint}\fi

\begin{abstract}
Given a fixed set of pending deletion requests, retraining from scratch after each request is prohibitive, so a prescribed request-wise policy processes them sequentially. The resulting terminal model can depend on their order. Rather than prescribing an ordering rule, we study the permutation objective induced by the fixed policy and ask when it admits simpler structure. We identify two independent reductions: position additivity represents the objective by request--position costs, reducing optimization to assignment and, with a shared positional profile, sorting; suffix localization removes dependence on the distant prefix while retaining interactions among the surviving requests. Under shared affine updates, we characterize the quadratic interactions that obstruct additivity, prove the reductions' independence, and show that suffix-conditioned assignment improves the approximation rate from $O(\rho^L)$ to $O(\rho^{2L})$. Experiments recover both structures in executed objectives. A controlled damped-Newton sweep shows that stronger contraction shifts the objective toward shorter, more suffix-specific dependence, while two full-network policies exhibit distinct positional and within-suffix structure. Structures identified from compact execution sets also predict unseen orders. These results frame deletion ordering as identifying the computational structure induced by the executed updates.
\end{abstract}

\section{Introduction}
\label{sec:introduction}
A deployed model is often required to remove specific training data after the
fact, and retraining from scratch on every request is prohibitive, so
approximate unlearning edits the trained model in place
\citep{guo2020certified,kurmanji2023towards,koloskova2025certified}.
When a prescribed request-wise policy processes a pending set, each request
acts on the model produced by its predecessors.  The same pending requests can
therefore lead to different terminal models under different orders; with the
request set, policy, and per-request budget fixed, the order is the decision
variable.

Existing work has shown that deletion order matters and has proposed schedules
based on uncertainty, difficulty, or forgettability
\citep{zhao2024what,kumar2026controlling}. These rules choose an order, but do
not characterize the optimization problem induced by the sequential updates.
The objective over the $k!$ schedules of $k$ pending requests may reduce to
independent request scores,
request--position costs, dependence on only the most recent requests, or none
of these. Unlike classical scheduling and training-order problems, where the
cost structure is prescribed
\citep{smith1956various,rajput2022permutation,li2025optimal},
here it is created by composing the deletion updates and evaluating the
terminal model.

\begin{figure}[t]
    \centering
    \includegraphics[width=\textwidth]{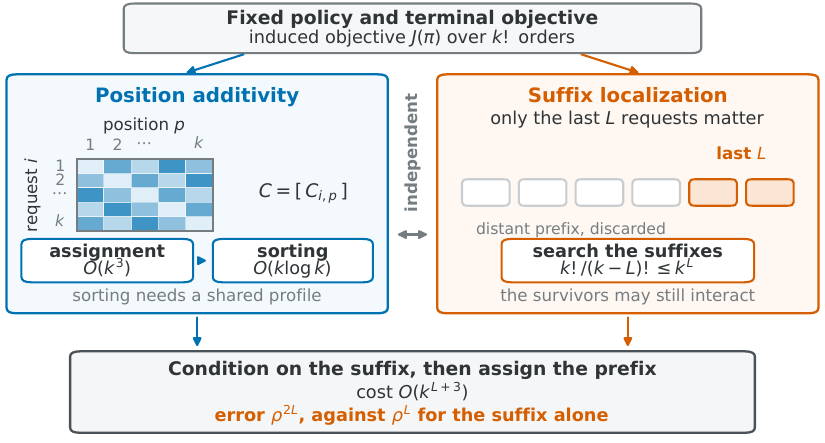}
    \caption{\textbf{Two independent routes to a simpler ordering problem.}
    Left, position additivity represents the objective by request--position
    costs.  Right, suffix localization discards the distant prefix.  Below,
    the two compose, at the rate Proposition~\ref{prop:common-suffix-memory}(c)
    gives for shared affine updates.}
    \label{fig:reduction-routes}
\end{figure}

We fix the deletion requests, update policy, and terminal objective, and vary
only their order, defining an objective $J(\pi)$ over permutations
$\pi\in S_k$, where $S_k$ is the set of all $k!$ request orders.  We study
two ways in which this objective can be simplified
(Figure~\ref{fig:reduction-routes}).
\emph{Position additivity} represents $J(\pi)$ by request--position costs,
reducing order selection to linear assignment and, under a shared positional
profile, to sorting.
\emph{Suffix localization} makes the distant prefix irrelevant once the final
$L$ requests are fixed, reducing the search to ordered suffixes.
The two reductions discard different information. Position additivity removes
cross-request interactions across positions, whereas suffix localization
removes dependence on the distant prefix. They can therefore hold separately
or together.

The update dynamics determine which reduction, if any, is available.
Even when request responses combine additively at the endpoint, a squared
terminal objective introduces pairwise request--position interactions, so
endpoint additivity need not imply position additivity.
Conversely, contractive updates suppress dependence on the distant prefix
without eliminating interactions among the surviving requests.
We characterize both mechanisms, prove that the two reductions are independent
and composable, and show that contraction can yield a useful approximation
without making exact optimization tractable.

We then test these structures in executed unlearning objectives.
Across frozen-head and full-network settings, the induced objectives occupy
different structural regimes. A deterministic damped-Newton intervention,
used here because its Jacobian is analytic, shows
that stronger contraction shifts dependence toward shorter, more
suffix-specific representations, while held-out studies show that structures
identified from compact execution sets transfer to unseen orders.
The cost of identifying a representation from executions is measured
separately from the cost of optimizing once that representation is available.

\paragraph{Contributions.}
We make three contributions:
\textbf{(i)} we formulate deletion ordering as an induced permutation objective
and identify position additivity and suffix localization as two independent
structural reductions;
\textbf{(ii)} we derive assignment, sorting, and finite-suffix reductions with
approximation guarantees, characterize their composition, and establish a
boundary for exact optimization under contraction; and
\textbf{(iii)} we connect these structures to executed frozen-head and
full-network dynamics and test their transfer beyond exhaustive small-$k$
tables.

\section{Related Work}
Approximate unlearning uses gradients, curvature, distillation, or parameter
saliency~\citep{guo2020certified,kurmanji2023towards,fan2024salun,
foster2024fast,koloskova2025certified}.  Reference proximity and behavioral
forgetting provide different evidence about an unlearning policy
~\citep{mavrothalassitis2025ascent}.  We use reference distance as a dense
terminal ordering objective and report behavioral gates and deletion
trajectories separately.
Guo et al. use a Newton correction, with loss perturbation for the general
case, whereas Koloskova et al. use noisy fine-tuning on retained data.  Our
deterministic damped-Newton rule is an intervention chosen for its known
Jacobian.

Zhao et al. compare one-shot unlearning with sequential processing of
difficulty-refined forget subsets, and Kumar et al. study noncommutativity and
uncertainty-based forget-set ordering~\citep{zhao2024what,
kumar2026controlling}.  These studies establish that request composition and
order affect sequential unlearning.  Our complementary question is which
permutation objective those executions induce and when that objective admits
sorting, assignment, or suffix reduction.  Request-level scoring corresponds to the
scalar-sorting specialization of this framework.

Adaptive and continual unlearning address deletion requests that arrive over
time, including validity under adaptively chosen requests and utility across
repeated language-model updates~\citep{gupta2021adaptive,gao2025continual}.
Our setting conditions on a finite pending batch and studies the objective over
its execution orders, isolating schedule dependence under a fixed sequential
policy.

Yu et al. study how the order of multi-stage training changes subsequent local
unlearning that uses forget-set gradients~\citep{yu2025impossibility}.  We hold
the trained starting state fixed and vary the later deletion history; common
deletion suffixes, rather than common training histories, define our memory
measure.  The changed history and the information available to the unlearning
rule are therefore different in the two settings.

Classical scheduling derives sorting rules from supplied job costs and
processing times, as in Smith's rule for weighted completion
time~\citep{smith1956various}.  Research on task and training order studies how
curricula, optimization stability, and function composition create history
dependence~\citep{li2025optimal,hardt2016train,
singh2023learning,cooper2023coordinating,kubo2024composition,
rajput2022permutation,lu2022general}.  Our tools draw on quadratic assignment
and weak-sum linearizability~\citep{wang2021rankone,punnen2013linear,
cela2014linearizable}.  In our setting, the costs are induced by a deletion
policy and terminal objective, and each permutation unit is a stateful request
that may contain many examples and internal updates.

Kubo et al. prove target-composition hardness for general increasing linear
maps~\citep{kubo2024composition}.  Our \textsc{Partition} construction instead
isolates the shared strict-contraction subclass: every request has linear part
$\rho I$ and is realized by a positive, non-overshooting gradient step, while
the dimension grows with the instance and $\rho$ approaches one.  The detailed
comparison is in Appendix~\ref{app:assignment-hardness}.

\section{Deletion Ordering as an Induced Computational Problem}
\label{sec:convex-setup}

Let the training set be the disjoint union of a final retain set $R$ and
deletion requests $F_1,\ldots,F_k$. Training on all data initializes the
execution state $w_0$. We fix an unlearning policy $\mathcal U$ and vary only
the request permutation. All stochastic choices are bound to request
identity, so changing the order changes composition rather than the random draw assigned
to a request.

Request $F_i$ induces a transition $U_i$, where the state $w$ includes both
the model and any policy state carried across requests. For
$\pi\in S_k$,
\begin{equation}
  w_\pi=U_{\pi_k}\circ\cdots\circ U_{\pi_1}(w_0).
\end{equation}

\paragraph{Terminal objective.}
Let $\Phi$ be a declared comparison map and $z_R$ a retained-data reference.
Define the state objective $j(w)=\|\Phi(w)-z_R\|_2^2$.  Our primary
permutation objective is
\begin{equation}
  J(\pi)=j(w_\pi)=\|\Phi(w_\pi)-z_R\|_2^2,
  \label{eq:jsq}
\end{equation}
with lower values preferred.  The comparison map lets one formulation cover
parameter-, output-, and representation-level terminal objectives. The
concrete instantiations appear in Section~\ref{sec:convex-experiments}.
We define
\[
J^\star=\min_{\pi\in S_k}J(\pi),\qquad
\Delta_J=\mathbb E_{\pi\sim\mathrm{Unif}(S_k)}J(\pi)-J^\star.
\]
Thus $\Delta_J$ is the improvement available over a uniformly random order.
We normalize structural errors only when $\Delta_J>0$.

\paragraph{Two structural reductions.}
For \emph{position additivity}, we approximate the executed objective $J$ by an
assignment-form surrogate
\begin{equation}
f(\pi)
=
c+\sum_{p=1}^k C_{\pi_p,p},
\label{eq:assignment-family}
\end{equation}
where $c\in\mathbb R$ and $C\in\mathbb R^{k\times k}$ are free coefficients, so
$C_{i,p}$ is the cost charged for serving request $i$ at position $p$.  Let
$\mathcal A$ denote the family of all such surrogates.  Exact position
additivity means that some $f\in\mathcal A$ satisfies $f(\pi)=J(\pi)$ for every
$\pi\in S_k$.

For \emph{suffix localization}, write $\pi=\sigma s$, where $s$ is an ordered
length-$L$ suffix and $\sigma$ its prefix. For example, for four requests
$w,x,y,z$, if $\pi=(w,x,y,z)$ and $L=2$, then
$\sigma=(w,x)$ and $s=(y,z)$. We define
\begin{equation}
W_L
=
\max_s
\left\{
\max_\sigma J(\sigma s)-\min_\sigma J(\sigma s)
\right\}.
\label{eq:suffix-width-definition}
\end{equation}
The width $W_L$ measures how much the objective can still vary with the
prefix after the final $L$ requests are fixed. Thus $W_L=0$ means that
$J$ depends only on the ordered length-$L$ suffix, while interactions within
that suffix may still be arbitrary.

\paragraph{Uniform structural error.}
For position additivity, define the normalized assignment error
\begin{equation}
D_{\mathcal A}
=
\frac{2}{\Delta_J}
\min_{f\in\mathcal A}
\max_{\pi\in S_k}|J(\pi)-f(\pi)|.
\label{eq:assignment-error}
\end{equation}
Thus $D_{\mathcal A}=0$ under exact position additivity.

For suffix localization, let $\mathcal S_L$ be the family of functions that
depend only on the ordered length-$L$ suffix.  The best constant for each fixed
suffix is the midpoint of its objective range, so
$\inf_{f\in\mathcal S_L}\|J-f\|_\infty=W_L/2$. We therefore define
\begin{equation}
D_{S,L}
=
\frac{W_L}{\Delta_J}.
\label{eq:suffix-error}
\end{equation}
Thus $D_{S,L}=0$ under exact length-$L$ suffix localization. Throughout,
\emph{width} denotes the raw range $W_L$, while \emph{error} denotes the
corresponding quantity normalized by $\Delta_J$.

The uniform norm gives a direct decision guarantee. If a surrogate $f$
satisfies $\|J-f\|_\infty\leq\varepsilon$ and
$\widehat\pi\in\arg\min_\pi f(\pi)$, then
\[
J(\widehat\pi)-J^\star\leq2\varepsilon,
\]
which we call the \emph{uniform regret bound}.
We therefore normalize uniform structural errors by $2/\Delta_J$. An error
below one is sufficient to guarantee that the selected order improves on the
uniformly random-order mean.

\section{Update Dynamics and Ordering Structure}
\label{sec:convex-theory}

A fixed unlearning policy induces a permutation objective through its update
dynamics. We begin with the simplest analyzable case, where all requests share
the same linear response, and ask when the resulting ordering problem reduces
below exhaustive search over $S_k$. Throughout this section, $w$ denotes the
representation in which terminal distance is measured, so
Equation~\ref{eq:jsq} has $\Phi=I$. Proposition~\ref{prop:common-suffix-memory}(a)
separately allows a Lipschitz nonlinear comparison map.

\subsection{Squared terminal distance creates interactions}

Suppose every request has the shared affine transition $U_i(w)=Mw+b_i$.
Repeated application gives the additive endpoint
\begin{equation}
  w_\pi=M^kw_0+\sum_{p=1}^k M^{k-p}b_{\pi_p}.
  \label{eq:shared-response}
\end{equation}
Had the terminal cost been linear, substituting
Equation~\ref{eq:shared-response} would preserve this additivity and give an
assignment objective $\mathrm{const}+\sum_p C_{\pi_p,p}$ with
$C_{i,p}=c^\top M^{k-p}b_i$.  The squared terminal objective instead
introduces pairwise interactions.  Expanding
Equation~\ref{eq:jsq} gives
\begin{equation}
 J(\pi)=c_0+\sum_p C_{\pi_p,p}
 +2\sum_{p<q}\mathcal K_{\pi_p,\pi_q,p,q},
 \label{eq:shared-quadratic}
\end{equation}
where $\mathcal K_{ijpq}=b_i^\top(M^{k-p})^\top M^{k-q}b_j$ and the unary
terms absorb the reference and diagonal contributions.  Thus endpoint
additivity does not imply position additivity of the induced objective.
Squared terminal distance couples request pairs with position pairs.  For
$M=\lambda I$, the norm term is a Koopmans--Beckmann quadratic assignment
problem with rank-one positional coefficient matrix
$P_{pq}=\lambda^{2k-p-q}$.  Linear assignment is therefore a special, rather
than generic, induced structure.  The full expansion, the Gram-matrix reading,
and the perturbation bounds that carry a surrogate guarantee over to bounded
deviations from the shared response are in
Appendix~\ref{app:shared-qap}.

\subsection{Position additivity yields assignment and sorting}

Equation~\ref{eq:shared-quadratic} locates the obstruction: the coupling
between a request pair and a position pair.  Ordering reduces to assignment
exactly when that coupling can be absorbed into unary request--position costs,
on the linearizable boundary characterized by~\citet{punnen2013linear}
and~\citet{cela2014linearizable}.  Exact linearization is a strong
requirement, so we instead measure distance from it.  Fix a position pair
$(p,q)$ with $p<q$.  The off-diagonal coefficients of
Equation~\ref{eq:shared-quadratic} at that pair form a request-pair
interaction matrix
\[
 T^{pq}:=\left[2\mathcal K_{ijpq}\right]_{i,j}\in\mathbb R^{k\times k},
\]
and this interaction is assignment-compatible when it separates into one term
for the request at position $p$ and one for the request at position $q$.  We
therefore decompose each entry as
\[
 T^{pq}_{ij}=a^{pq}_i+\beta^{pq}_j+\Xi^{pq}_{ij},\qquad i\ne j,
\]
where the first two terms are absorbed into request--position costs and $\Xi$
is the remaining non-additive interaction.  Collecting the additive terms over
all position pairs yields a projected assignment matrix $\widehat C$.

Assignment collapses further to sorting when its costs share a common
monotone positional profile,
\[
 C_{i,p}=a_i+s_ih_p.
\]
For any fixed monotone profile $h$, the rearrangement inequality makes sorting
the request scores $s_i$ optimal.  To quantify approximate collapse with a
simple closed-form certificate,
Proposition~\ref{prop:slice-additive-certificate}(b) specializes this profile
to the centered linear choice $q_p=p-(k+1)/2$ and measures the residual from
that structure.

\begin{proposition}[Position reduction]
\label{prop:slice-additive-certificate}
(a) Suppose the executed objective differs uniformly from the shared quadratic
surrogate by at most $\varepsilon_{\rm surr}$.  Let $\widehat\pi$ minimize the
projected assignment matrix and set
$\|\Xi\|_F^2=\sum_{p<q}\sum_{i\ne j}(\Xi^{pq}_{ij})^2$.  Then
\begin{equation}
 J(\widehat\pi)-\min_\pi J(\pi)
 \leq 2\varepsilon_{\rm surr}
 +2\sqrt{\binom{k}{2}}\,\|\Xi\|_F.
 \label{eq:slice-additive-regret}
\end{equation}
(b) Suppose instead that an assignment objective with cost matrix $C$
approximates $J$ uniformly within $\varepsilon_{\mathcal A}$.  Every
permutation pays a request-only and a position-only effect once, so
double-center, $\bar C=HCH$ with $H=I-\tfrac1k\mathbf 1\mathbf 1^\top$.  For
the centered linear profile $q_p=p-(k+1)/2$, write
\[
 \bar C=s q^\top+\bar\Xi_q
\]
as the projection of $\bar C$ onto the corresponding sorting structure, and
let $\pi_s$ sort $s$.  Then
\begin{equation}
 J(\pi_s)-\min_\pi J(\pi)
 \leq 2\varepsilon_{\mathcal A}+\sqrt{2k}\,\|\bar\Xi_q\|_F.
 \label{eq:q-regret}
\end{equation}
\end{proposition}
The two parts chain.  The assignment matrix that part (a) projects out satisfies
$\varepsilon_{\mathcal A}\leq\varepsilon_{\rm surr}
+\sqrt{\binom{k}{2}}\,\|\Xi\|_F$, so substituting this into part (b) bounds
the sorted order's regret by the two residuals together.  A small interaction residual therefore certifies
approximate position additivity, while a small linear-profile residual
certifies that the corresponding scalar sorting rule suffices.
The retained-gradient construction is in
Appendix~\ref{app:shared-qap}, the linearizability details in
Appendix~\ref{app:assignment-hardness}, and both projection proofs, together
with the scope of the slice projection, in
Appendix~\ref{app:auxiliary-projections}.

\subsection{Contraction localizes prefix dependence and composes with assignment}

\paragraph{Memory reduction.}
Contraction provides a different simplification from additivity.  A common
contractive suffix attenuates differences between states produced by different
prefixes.  If the terminal objective is Lipschitz on the resulting states, its
dependence on the distant history contracts accordingly, while interactions
within the suffix remain unrestricted.
Proposition~\ref{prop:common-suffix-memory}(a,b) quantifies this, first for
Lipschitz updates and then for the affine case.

\paragraph{Composition.}
Simply truncating the prefix also discards its interactions with the retained
suffix, leaving an $O(\rho^L)$ error term.  A sharper construction fixes the
ordered suffix and absorbs these prefix--suffix interactions into unary
request--position costs for the remaining requests.  The only discarded terms
are then prefix--prefix interactions, whose two factors have each been
attenuated by at least $L$ subsequent updates.  This yields the
$O(\rho^{2L})$ bound in Proposition~\ref{prop:common-suffix-memory}(c).

\begin{proposition}[Memory reduction and composition]
\label{prop:common-suffix-memory}
\label{prop:nonlinear-common-suffix}
\label{prop:finite-memory-ordering}
(a) Fix an ordered length-$L$ suffix $s$.  Let $\mathcal X_s$ be its reachable
pre-suffix states with diameter $d_s$, and suppose the suffix map $U_s$, the
composition of the final $L$ updates, is $\kappa_s$-Lipschitz on
$\mathcal X_s$.  If $j$ is $L_{j,s}$-Lipschitz on $U_s(\mathcal X_s)$, then
\begin{equation}
 \max_{u,v\in\mathcal X_s}|j(U_s(u))-j(U_s(v))|
 \leq L_{j,s}\kappa_s d_s.
 \label{eq:nonlinear-common-suffix-bound}
\end{equation}
For the squared objective of Equation~\ref{eq:jsq} with $L_\Phi$-Lipschitz
$\Phi$, one may take $L_{j,s}=2R_sL_\Phi$ with
$R_s=\sup_{w\in U_s(\mathcal X_s)}\|\Phi(w)-z_R\|$.
(b) Under the shared affine transition with $\|M\|\leq\rho<1$ and
$\max_i\|b_i\|\leq B$, retaining only the final $L$ response terms approximates
$J$ uniformly over orders to $O(\rho^L)$, so the minimizer of the truncated
objective has regret of the same order.
(c) Under the hypotheses of (b), write $\pi=\sigma s$ and define the pointwise
surrogate
$\widehat J_L(\sigma s)=\|r_s\|^2+\sum_{p\leq k-L}C^{(s)}_{\sigma_p,p}$,
where $r_s$ is the residual fixed by $s$ and $C^{(s)}$ contains the
suffix-conditioned unary prefix costs.  Then
\begin{equation}
 \sup_\pi|J(\pi)-\widehat J_L(\pi)|
 \leq\frac{B^2\rho^{2L}}{(1-\rho)^2},
 \label{eq:suffix-conditioned-error}
\end{equation}
so its minimizer has regret at most $2B^2\rho^{2L}/(1-\rho)^2$.  Minimizing
$\widehat J_L$ enumerates the ordered suffixes and solves one assignment problem
per suffix, at cost $O(k^{L+3})$.
\end{proposition}
At a fixed suffix depth, conditioning on the suffix before assigning the prefix
therefore sharpens the affine rate from $\rho^{L}$ to $\rho^{2L}$.  Part (a),
separately, establishes suffix localization with no additivity assumption on
the surviving requests.  For fixed
contraction, uniformly bounded response constants, and a last-position
decision signal that does not vanish, the depth a fixed tolerance requires is
$L=O(\log(1/\tau))$, independent of $k$.
Direct truncation to depth $2L$ has the same affine rate at optimization cost
$O(k^{2L})$, so the conditioned construction has a smaller exponent in $k$
only when $L>3$.  These costs assume that the structural representation is
already available; constructing or identifying it is accounted for separately.
The non-asymptotic constants behind (b), the Lipschitz constants for
gradient-descent updates, the proofs, the queue-length result, and the
contrasting refinement regime are in Appendix~\ref{app:memory-bounds}.

\subsection{Independence and exact complexity}

\paragraph{Independence.}
Position additivity and suffix localization capture distinct properties of the
induced permutation objective, and neither subsumes the other.  A nilpotent
contraction yields exact suffix localization while retaining non-additive
interactions, whereas an exact assignment objective can retain positive suffix
width.

\begin{proposition}[Independence]
\label{prop:memory-additivity-separation}
Even under shared strict-contraction updates, position additivity and suffix
localization do not imply one another, and each separation can cross the
sufficient threshold of Section~\ref{sec:convex-setup}.
\end{proposition}
Explicit low-dimensional witnesses, the queue lengths and suffix depths they
cover, and their exact constants are in Appendix~\ref{app:memory-bounds}.

\paragraph{Exact complexity.}
The approximation structure induced by contraction does not determine the
complexity of exact optimization.  Even a positive, non-overshooting gradient
step can induce an NP-hard ordering problem.
\begin{proposition}[Hardness under arbitrarily weak strict contraction]
\label{prop:shared-response-hardness}
Deciding whether an order attains a given terminal objective is NP-complete
even when every request is one positive, non-overshooting gradient step on a
strictly convex quadratic, and hence a strict contraction with factor
$\rho<1$ that approaches one as the instance grows.
\end{proposition}
The weak NP-hardness reduction from \textsc{Partition} has $\rho\to1$ as the
instance grows.  It separates strict contraction from exact tractability but
leaves the complexity for $\rho$ bounded away from one open.  The construction,
proof, and fixed-contraction approximation are in
Appendix~\ref{app:assignment-hardness}.

\section{Empirical Structure of Executed Ordering Objectives}
\label{sec:convex-experiments}

We ask which structural forms executed objectives take, which update
dynamics produce them, and whether the structure can be identified from
limited executions and transferred to unseen orders.

\noindent\textbf{Settings.}
We study frozen-head and full-network unlearning.  Frozen heads include ridge
and softmax classifiers on CIFAR-10 features from an ImageNet-pretrained
ResNet-18~\citep{krizhevsky2009learning,he2016deep,deng2009imagenet}, and ridge
classifiers on AG News and 20 Newsgroups embeddings from
DistilBERT~\citep{zhang2015character,lang1995newsweeder,sanh2019distilbert}.
Full-network experiments delete CIFAR-100 superclasses from a ResNet-18 using
NegGrad+~\citep{kurmanji2023towards} and SalUn~\citep{fan2024salun}.  Queue
lengths range from four to fifteen; RGD, SMX, NG+, SalUn, and RNewt abbreviate
ridge gradient descent, softmax, NegGrad+, SalUn, and ridge Newton, with
\emph{-ext} denoting a queue extension.  Frozen heads use squared
parameter distance.  The full-network objective averages, over a fixed
2,000-example test audit set, squared Euclidean distance between centered
20-superclass-logit vectors and a fixed two-run retain-only reference.  It sums
coordinates within each example; update and audit data are disjoint.

\noindent\textbf{Protocol.}
Small queues are enumerated exhaustively, whereas longer-queue studies use
disjoint identification and evaluation orders.  To distinguish suffix-specific
dependence from generic low-dimensional positional structure, we compare
$D_{S,L}$ with equal-capacity prefix and middle controls $D_{P,L}$ and
$D_{M,L}$, defined by conditioning on the first $L$ positions or on a fixed
contiguous middle block.  Unless suffix depth is under study, $L=2$ and we
write $D_S$, $D_P$, and $D_M$.  All structural errors are normalized by the
random-order gap $\Delta_J$.  Full execution protocols and validity criteria
are in Appendix~\ref{app:protocols}.

\noindent\textbf{Objective anchoring.}
Reference distance compares completed sequences; utility and forgetting remain
separate.  On complete full-network surfaces it tracks retained more closely
than deleted-group accuracy (Appendix~\ref{app:behavioral-anchoring}).  Two
fixed reference-free objectives on three auxiliary replays retain assignment
and suffix-$2$ while matched controls fail (Appendix~\ref{app:reference-free});
this retrospective result is not a prospective pipeline.

\subsection{Executed objectives exhibit different structural forms}

\begin{figure}[t]
\centering
\begin{minipage}[t]{0.52\textwidth}
  \vspace{0pt}
  \centering
  \textbf{(a)} Structural regimes\par\smallskip
  \includegraphics[width=\linewidth]{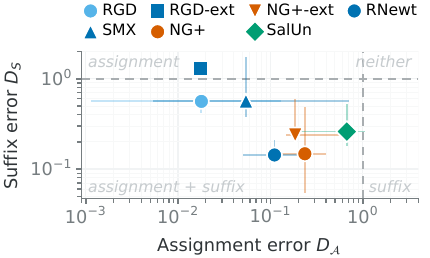}
\end{minipage}%
\hspace{0.02\textwidth}%
\begin{minipage}[t]{0.44\textwidth}
  \vspace{0pt}
  \centering
  \textbf{(b)} Response-level reconstruction\par\smallskip
  \includegraphics[width=\linewidth]{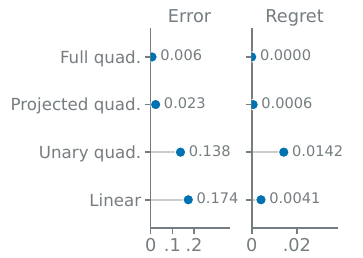}
\end{minipage}
\caption{\textbf{Structural regimes and response-level reconstruction.}
\textbf{(a)} Assignment and suffix errors across executed settings.  Points are
medians across instances, whiskers the observed ranges, and the dashed lines
mark $D=1$.  Quadrant labels refer to the median point, and \emph{-ext} marks a
queue-extension setting.
\textbf{(b)} Reconstruction error $2\varepsilon/\Delta_J$ and selection regret
for four singleton-response constructions, over $45$ complete $k=4$
fresh-decision surfaces.}
\label{fig:structural-axes}
\end{figure}

Executed objectives occupy different regimes
(Figure~\ref{fig:structural-axes}a).  The exhaustive ridge extension is
assignment-only.  Complete full-network surfaces support suffix localization;
NegGrad+ also \mbox{supports} assignment throughout, while SalUn is more heterogeneous
on that axis.  Proposition~\ref{prop:memory-additivity-separation} supplies the
complementary suffix-only separation.

\noindent\textbf{Position axis.}
A learned shared monotone profile brings NegGrad+ close to unrestricted
assignment, while SalUn remains more heterogeneous; a fixed linear profile can
therefore understate proximity to sorting.

\noindent\textbf{Suffix axis.}
A low suffix error alone does not establish suffix-specific dependence.  For
frozen-head gradient objectives, an equal-capacity prefix fit is comparably
accurate.  On full-network objectives, matched prefix and middle controls
separate the suffix structure.  On the complete $k=4$ surfaces, NegGrad+ is
localized to the final request, whereas SalUn generally requires the final two
(Figure~\ref{fig:mechanism}b); its additive two-position suffix is also looser,
showing residual interactions (Appendix~\ref{app:family-separation}).

\FloatBarrier
\subsection{Update dynamics shape the induced structure}

\begin{figure}[t]
  \centering
  \includegraphics[width=0.96\textwidth]{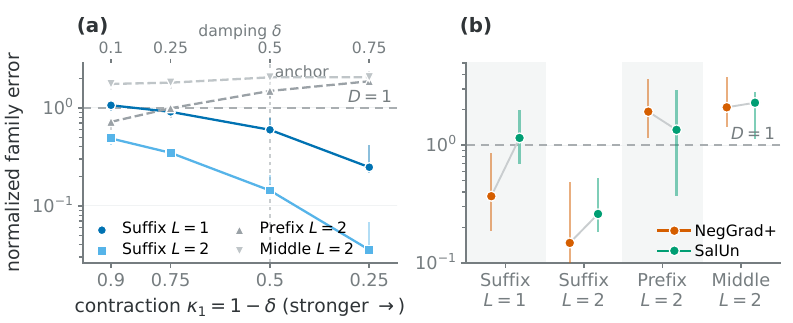}
  \caption{\textbf{Update dynamics and positional memory.}  Points show
  medians across instances and whiskers the observed ranges, and the dashed line
  marks the normalized-error threshold $D=1$.  \textbf{(a)} Damped-Newton sweep,
  plotted against the analytic contraction $\kappa_1=1-\delta$, with the damping
  that sets it on the upper axis.  \textbf{(b)} NegGrad+ and SalUn compared across the suffix families
  and their equal-capacity prefix and middle controls.}
  \label{fig:mechanism}
\end{figure}

\noindent\textbf{Quadratic interactions.}
Using the same singleton responses, the quadratic terminal score reconstructs
the executed objective substantially more accurately than either its linear
part or unary quadratic corrections carrying no interaction term
(Figure~\ref{fig:structural-axes}b).  Projecting those interactions onto
request--position costs retains most of that accuracy and produces a tractable
assignment objective.  All constructions use no terminal-order
observations, and the pattern holds across vision and text representations.  A
matched step-size intervention independently recovers the predicted
second-order scaling, with a fitted slope of $1.97$ against a predicted $2$.

\noindent\textbf{Controlled contraction.}
Varying Newton damping shifts the objective toward shorter, more
suffix-specific representations (Figure~\ref{fig:mechanism}a).  For damped
Newton the retained Hessian cancels, leaving a Jacobian of $(1-\delta)I$ and
hence known contraction $\kappa_1=1-\delta$.  On all five held-out instances,
both suffix errors fall as contraction strengthens; the equal-capacity prefix
error rises monotonically, and the middle control rises on four of five.
Assignment is non-monotone, so the intervention changes where positional
memory resides rather than simplifying the objective uniformly.

\noindent\textbf{Beyond affine updates.}
Nonlinear full-network policies show the same localization.  After a common
two-request suffix, histories from different prefixes retain only $6.9\%$ of
their median objective difference under NegGrad+ and $10.2\%$ under SalUn.
At $k=4$, effective suffix depth is one request for NegGrad+ and generally two
for SalUn (Figure~\ref{fig:mechanism}b).  These are finite-execution
observations; the
causal claim rests on the controlled damping intervention above
(Appendix~\ref{app:suffix-mechanism}).

\noindent\textbf{Composition.}
On the three complete $k=5$ NegGrad+ surfaces, we compare rank-50 assignment
models conditioned on the first, third, or final request.  Four design-only
folds hold out 30 cyclic-relabeling-balanced orders each; every 90-order
training design has full rank, and fitting uses training-fold normalization.
\begin{table}[H]
\centering
\footnotesize\begin{tabular}{lrrr}
\toprule
Instance & Prefix-cond. & Middle-cond. & Suffix-cond. \\
\midrule
seed0 & $0.477$ $[0.433,0.522]$ & $0.370$ $[0.347,0.501]$ & $0.308$ $[0.285,0.323]$ \\
seed1 & $1.579$ $[1.325,1.941]$ & $1.547$ $[1.215,2.053]$ & $0.724$ $[0.630,0.990]$ \\
seed2 & $0.381$ $[0.365,0.409]$ & $0.349$ $[0.299,0.357]$ & $0.170$ $[0.157,0.190]$ \\
\bottomrule
\end{tabular}%

\caption{\textbf{Matched-rank position conditioning.}  Entries are the median
held-out $D^{\rm test}$ with observed $[\min,\max]$ over four design-only folds.}
\label{tab:conditioned-position-holdout}
\vspace{-5pt}
\end{table}
Final-request conditioning wins every instance--fold comparison, separating
suffix specificity from capacity.  The folds are not independent instances,
and this one-depth test does not test the nonlinear $\rho^{2L}$ exponent.

\subsection{Identified structure transfers to unseen orders}

\begin{figure}[t]
\centering
\begin{minipage}[t]{0.54\textwidth}
  \vspace{0pt}
  \centering
  \textbf{(a)} Ridge identification and transfer\par\smallskip
  \small
  \begin{tabular}{@{}rlrr@{}}
\toprule
Requests & Family & Fit executions & Held-out error \\
\midrule
8 & Sorting & 8 & $0.068$ \\
8 & Assignment & 50 & $0.055$ \\
\addlinespace[2pt]
15 & Sorting & 15 & $0.107$ \\
15 & Assignment & 197 & $0.110$ \\
\bottomrule
\end{tabular}%

\end{minipage}%
\hspace{0.02\textwidth}%
\begin{minipage}[t]{0.43\textwidth}
  \vspace{0pt}
  \centering
  \textbf{(b)} Full-network held-out error\par\smallskip
  \includegraphics[width=\linewidth]{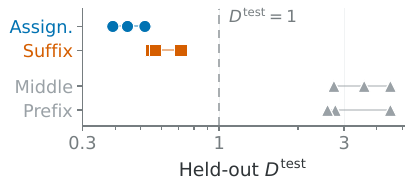}
\end{minipage}
\caption{\textbf{Structural identification transfers to unseen orders.}
\textbf{(a)} Sorting and assignment fits identified from compact execution sets
and scored on orders held out of the fit, exhaustively at eight requests and on
a disjoint $500$-order sample at fifteen.  \textbf{(b)} Held-out structural
error $D^{\rm test}$ for the
full-network prefix, middle, suffix, and assignment families, one point per
instance, with the dashed line at $D^{\rm test}=1$.}
\label{fig:fullnet-identification}
\end{figure}

The ridge identification orders are fixed from the family design matrix before
objective values are observed.  With eight identification executions, sorting
recovers the unique optimum among all 40,320 orders in each of the three
$k=8$ ridge instances.  Assignment fits more tightly and recovers the same
optima, but requires 50 executions (Figure~\ref{fig:fullnet-identification}a).
At $k=15$, both families remain accurate on held-out orders.  Newly executed
minimizers of the fitted assignment objectives outperform every identification
and held-out order in all three instances
(Appendix~\ref{app:quadratic-ablation}), although their global ranks remain
unknown.

For full-network CIFAR-100, each instance uses 112 identification and 336
disjoint held-out orders.  Assignment and suffix fits remain below threshold
on held-out orders, whereas matched prefix and middle controls exceed it
(Figure~\ref{fig:fullnet-identification}b).  Executed unseen candidates pass
the original screens, with mixed realized ranks
(Appendix~\ref{app:fullnet-candidates}); this tests structural extrapolation,
not search efficiency.

\section{Discussion and Conclusion}
\label{sec:discussion}

Deletion ordering need not admit a universal rule because its computational
structure is induced jointly by the update dynamics and terminal objective.
Position additivity and suffix localization expose two distinct simplifications:
one removes cross-request positional interactions, while the other removes
distant history without eliminating interactions among recent requests.  This
distinction explains why contraction can shorten memory without making the
objective additive, and why suffix conditioning can compose with assignment.
This also separates approximate from exact tractability: contraction can make
recent-history reductions accurate, while strict contraction alone does not
guarantee easy exact optimization.
Across the executed settings, these structural regimes remain distinguishable
under matched controls and can transfer to orders outside the identification
set.

The resulting perspective is to treat deletion ordering as a structure
identification problem before choosing a solver.  A shared positional profile
supports sorting, request--position costs support assignment, and localized
history reduces the search to recent suffixes.  Which reduction is appropriate
therefore depends on what information the executed dynamics preserve in the
terminal objective.  A richer family is not automatically preferable: on the
complete ridge surfaces, assignment fits more tightly, but sorting reaches the
same optimum from fewer identification executions.  Useful reductions must
also remain predictive beyond their identification executions.

\label{endofmain}

\clearpage
\section*{Reproducibility statement}

\ifdefined\preprint
A public repository will provide the frozen protocols, the raw
\else
The anonymous supplementary repository contains the frozen protocols, the raw
\fi
execution outputs, and the scripts needed to regenerate every reported table
and figure and to verify the numerical claims.  Analyses that use outcomes
retrospectively are identified at first use.  Held-out and pre-outcome
constructions are versioned in the accompanying protocols, together with the
partitions and checkpoints each run is bound to.

\section*{Use of AI assistants}

AI assistants were used for research discussion, code development and
debugging, experimental analysis, and manuscript editing.  The authors
verified all code, results, and claims and take responsibility for the content.

\bibliography{iclr2027_conference}
\bibliographystyle{iclr2027_conference}

\clearpage
\appendix
\label{startofappendix}

\section{Proofs and theoretical details}
\label{app:convex-proofs}

\subsection{Exact and approximate structure under shared responses}

\paragraph{Exact assignment under retained-gradient updates.}
A retained-gradient least-squares instance realizes exact position additivity.
Let
\[
\mathcal L_R(w)=\tfrac12h_R\|w\|^2-b_R^\top w,
\qquad
\mathcal L_i(w)=\tfrac12h\|w\|^2-b_i^\top w,
\]
with $h_R,h>0$ and $b_i^\top b_j=0$ for $i\ne j$, take the terminal reference
to be the retained-data minimizer $z_R=h_R^{-1}b_R$, and starting at the
full-data minimizer take one fixed-size gradient step on the current retained
loss after each deletion.  Expanding the recurrence gives
$w_\pi=\gamma_Rb_R+\sum_p a_pb_{\pi_p}$ with $a_p$ depending only on position,
so $w_\pi-z_R=v+\sum_pa_pb_{\pi_p}$ for $v=(\gamma_R-h_R^{-1})b_R$, and
orthogonality removes the cross-request terms to leave
\[
  J(\pi)=c_0+\sum_{p=1}^k C_{\pi_p,p},
  \qquad C_{i,p}=a_p^2\|b_i\|^2+2a_pv^\top b_i.
\]
Linear assignment returns a global optimum.
Writing $r_t=1-\eta\{h_R+(k-t)h\}$ and $P_t=\prod_{u=t}^kr_u$ with an empty
product equal to one, the coefficients are
$a_p=P_1/(h_R+kh)+\eta\sum_{t<p}P_{t+1}$ and
$\gamma_R=P_1/(h_R+kh)+\eta\sum_{t\leq k}P_{t+1}$.  These quadratics are
ordinary least-squares blocks.  Taking $X_i=\sqrt hI$ and $y_i=b_i/\sqrt h$
gives $X_i^\top X_i=hI$ and $X_i^\top y_i=b_i$, with the retain block
analogous.

\paragraph{Interactions under general shared responses.}
\label{app:shared-qap}
Substituting Equation~\ref{eq:shared-response} into the squared terminal
objective gives Equation~\ref{eq:shared-quadratic}.  Specifically, let
$\Psi_p=M^{k-p}$, $r=M^kw_0-z_R$,
\[
 c_0=\|r\|^2,\qquad
 C_{i,p}=2r^\top \Psi_pb_i+b_i^\top \Psi_p^\top \Psi_pb_i,
 \qquad \mathcal K_{ijpq}=b_i^\top \Psi_p^\top \Psi_qb_j.
\]
The squared terminal objective therefore couples request pairs $(i,j)$ with
position pairs $(p,q)$ through the four-index tensor $\mathcal K$.  For the
scalar response $M=\rho I$ it factors as
$\mathcal K_{ijpq}=\rho^{\,2k-p-q}\langle b_i,b_j\rangle$, separating a
rank-one positional matrix from the request Gram matrix, which is the form
Appendix~\ref{app:assignment-hardness} uses to discuss linearizability.

\paragraph{Robustness to response mismatch.}
The following uniform bounds quantify the additional surrogate error when the
executed responses deviate from the shared-response model.  They require the
stated bounds to hold uniformly over reachable states, so estimating the same
quantities only along realized trajectories gives a finite-execution
diagnostic rather than a uniform guarantee.

\emph{Linear mismatch.}  Let $U_i(w)=(M+E_i)w+b_i$ and $\bar U_i(w)=Mw+b_i$,
with $\|M\|\leq a$, $\|E_i\|\leq\delta$, $\|b_i\|\leq B$, and
$\bar a=a+\delta$.  Writing $\widetilde M_t=M+E_{\pi_t}$ for the executed linear
parts, the telescoping identity
$\widetilde M_m\cdots\widetilde M_1-M^m
=\sum_\ell\widetilde M_m\cdots\widetilde M_{\ell+1}E_{\pi_\ell}M^{\ell-1}$
with submultiplicativity gives
$\|\widetilde M_m\cdots\widetilde M_1-M^m\|\leq m\delta\bar a^{\,m-1}$.
Applying it to the initial state, which is multiplied by a product of length
$k$, and to a displacement inserted at position $p$, which is multiplied by the
$k-p$ matrices after it, gives
\begin{equation}
 \|w_\pi-\bar w_\pi\|\leq
 \Gamma_k:=k\delta\bar a^{\,k-1}\|w_0\|
 +B\delta\sum_{m=1}^{k-1}m\bar a^{\,m-1}.
 \label{eq:perturbed-shared-endpoint}
\end{equation}
Expanding $\|w_\pi-z_R\|^2-\|\bar w_\pi-z_R\|^2$ in $e_\pi=w_\pi-\bar w_\pi$
then bounds the terminal objectives uniformly by $2R_k\Gamma_k+\Gamma_k^2$,
where $R_k=\max_\pi\|\bar w_\pi-z_R\|\leq\|M^kw_0-z_R\|+B\sum_{r<k}a^r$.

\emph{Nonlinear mismatch.}  Fix a common initial state, write $v=w-w_0$, and
suppose that, uniformly over every reachable state $v$, deletion history $S$,
and next request $i$,
\begin{equation}
 U_{S,i}(w_0+v)-w_0
 =Mv+b_i+E_{S,i}v+d_{S,i}+r_{S,i}(v),
 \label{eq:nonlinear-response-decomposition}
\end{equation}
with $\|M\|\leq a$, $\|E_{S,i}\|\leq\delta$, $\|d_{S,i}\|\leq\zeta$, and
$\|r_{S,i}(v)\|\leq K\|v\|^2/2$.  If $\|v_t\|\leq Q_t$,
subtracting the shared surrogate recurrence $\bar v_{t+1}=M\bar v_t+
b_{\pi_{t+1}}$ and inducting on
\begin{equation}
 \Gamma_{t+1}=a\Gamma_t+\delta Q_t+\zeta+\frac K2Q_t^2,
 \qquad \Gamma_0=0,
 \label{eq:nonlinear-response-recurrence}
\end{equation}
gives $\|v_t-\bar v_t\|\leq\Gamma_t$, and the same expansion as above bounds
the terminal objectives by $2\bar R_k\Gamma_k+\Gamma_k^2$ for
$\bar R_k=\max_\pi\|w_0+\bar v_k-z_R\|$.

Write $J_{\rm aff}$ for the shared-response objective and let
$\varepsilon_{\rm dyn}$ be the bound just obtained, so
$\|J-J_{\rm aff}\|_\infty\leq\varepsilon_{\rm dyn}$.  If a surrogate $\widehat J$
satisfies $\|\widehat J-J_{\rm aff}\|_\infty\leq\varepsilon_{\rm surr}$, then
$\|J-\widehat J\|_\infty\leq\varepsilon_{\rm dyn}+\varepsilon_{\rm surr}$, so
the uniform regret bound gives any minimizer of $\widehat J$ regret at most
$2(\varepsilon_{\rm dyn}+\varepsilon_{\rm surr})$.

\subsection{Assignment reductions and exact hardness}
\label{app:assignment-hardness}

\paragraph{Scalar-response linearizability.}
For $M=\rho I$, let $\alpha_p=\rho^{k-p}$,
$P=\alpha\alpha^\top$, and $G_{ij}=\langle b_i,b_j\rangle$.  Equivalently
grouping the diagonal terms back into the quadratic term, the norm-generated
part has the Koopmans--Beckmann form
\[
 \sum_{p,q}P_{pq}G_{\pi_p,\pi_q}.
\]
The rank-one qualifier here applies to the positional coefficient matrix
$P=\alpha\alpha^\top$.  The request Gram matrix $G$ may have higher rank and
the request vectors may have arbitrary dimension.  Related work studies QAPs
with rank-one cost matrices~\citep{wang2021rankone}.  Here we state
the
restriction directly in terms of $P$ and $G$, and what the hardness
construction adds is the realization of such an instance by positive,
non-overshooting gradient steps.  Linearizability is then a known
characterization rather than something we establish.  A symmetric
Koopmans--Beckmann QAP is linearizable as a linear assignment objective if and
only if at least one coefficient matrix is weak-sum, meaning $H_{ij}=u_i+u_j$
off diagonal~\citep{punnen2013linear,cela2014linearizable}.  What is specific
to our instances is where that condition fails.  For the position product
matrix,
\[
 P_{pq}+P_{rs}-P_{pr}-P_{qs}
 =(\alpha_p-\alpha_s)(\alpha_q-\alpha_r).
\]
Four distinct geometric position weights therefore violate the condition.
The same criterion identifies the exceptional request geometries through
their Gram matrix $G$.

Proposition~\ref{prop:slice-additive-certificate} measures distance from
assignment without assuming a scalar response, and applies here too.  Using $T^{pq}_{ij}=2\mathcal K_{ijpq}$, independently
least-squares project every off-diagonal slice onto
$a^{pq}_i+\beta^{pq}_j$.  Define
\[
 \widehat C_{i,p}=C_{i,p}
 +\sum_{q>p}a^{pq}_i+\sum_{q<p}\beta^{qp}_i.
\]
Then, up to the order-independent constant,
\[
 J(\pi)-\sum_p\widehat C_{\pi_p,p}
 =\sum_{p<q}\Xi^{pq}_{\pi_p,\pi_q}.
\]
Every permutation selects exactly $\binom{k}{2}$ residual entries, hence
\[
 \left|\sum_{p<q}\Xi^{pq}_{\pi_p,\pi_q}\right|
 \leq\sqrt{\binom{k}{2}}
 \left(\sum_{p<q}\sum_{i\ne j}(\Xi^{pq}_{ij})^2\right)^{1/2}.
\]
Combining this uniform error with $\varepsilon_{\rm surr}$ and applying
the uniform regret bound proves the proposition.  More generally, for
predicted responses $e_{i,p}$ and a quadratic terminal score
$\langle g,\delta\rangle+\tfrac12\langle\delta,H\delta\rangle$, use
\[
 C_{i,p}=\langle g,e_{i,p}\rangle
 +\tfrac12\langle e_{i,p},He_{i,p}\rangle,
 \qquad T^{pq}_{ij}=\langle e_{i,p},He_{j,q}\rangle.
\]
This is the form used by the projected retained-loss realization.  Slice
additivity is only sufficient.  Cancellations across position pairs can make a
tensor linearizable even when this residual is nonzero, and a large bound does
not exclude another tractable representation.

\paragraph{NP-completeness under positive small steps.}
To prove Proposition~\ref{prop:shared-response-hardness}, reduce from
Partition~\citep{gareyjohnson1979computers}.  Given positive integers
$a_1,\ldots,a_N$, set $s_i=2a_i$, $k=2N$, and $S=\sum_i s_i$.  Let
\[
 \mathcal C=\{-(2N-1),\ldots,-3,-1,1,3,\ldots,2N-1\}
\]
be the centered odd position weights.  In dimension $N+1$, with standard
basis $e_0,e_1,\ldots,e_N$, create two requests for each item:
\[
 b_{X_i}=\Omega e_i+s_ie_0,\qquad
 b_{Y_i}=-\Omega e_i-s_ie_0,\qquad \Omega=4NS+1.
\]
Their sum is zero.  If $X_i$ and $Y_i$ receive weights $c_{X_i}$ and
$c_{Y_i}$, write $d_i=c_{X_i}-c_{Y_i}$.  For
$v_\pi=\sum_pc_pb_{\pi_p}$,
\begin{equation}
 \|v_\pi\|^2
 =\Omega^2\sum_{i=1}^N d_i^2
  +\left(\sum_{i=1}^Ns_id_i\right)^2.
 \label{eq:positive-hardness-centered-objective}
\end{equation}

Every $d_i$ is a nonzero even integer, so $\sum_id_i^2\geq4N$ with equality
exactly when $|d_i|=2$.  Weights in $\mathcal C$ differing by two form a path
on $2N$ vertices whose perfect matching is unique by endpoint forcing, so
equality means adjacent weights are paired and $d_i=2\sigma_i$ with
$\sigma_i\in\{-1,1\}$.  Which pair goes to which item does not enter the
objective.  A proper pairing therefore has objective
$4N\Omega^2+4(\sum_i\sigma_is_i)^2\leq4N\Omega^2+4S^2$, while an improper one has some
$|d_i|\geq4$ and anchor term at least $(4N+12)\Omega^2$.  Since $12\Omega^2>4S^2$, every
global optimum is properly paired and
\begin{equation}
 \min_\pi\|v_\pi\|^2
 =4N\Omega^2+4\min_{\sigma\in\{-1,1\}^N}
   \left(\sum_i\sigma_i s_i\right)^2.
 \label{eq:positive-hardness-pairing-identity}
\end{equation}

It remains to embed these centered linear weights in a positive geometric
response.  Set $B=\Omega+S$ and
\[
 \eta=\frac{1}{100B^2k^6},
 \qquad \rho=1-\eta.
\]
For position $p=1,\ldots,k$, let
$c'_p=2(k-p)-(k-1)=k+1-2p$, which traverses $\mathcal C$.  Because
$\sum_i(b_{X_i}+b_{Y_i})=0$, the constant term in
$(1-\eta)^{k-p}$ cancels.  Taylor's theorem, with
$(1-\eta)^j=1-j\eta+r_j$ and $0\leq r_j\leq j(j-1)\eta^2/2$, writes the exact
endpoint under $U_i(w)=\rho w+b_i$ as $w_\pi=-\tfrac\eta2v_\pi+\xi_\pi$.
Since $\sum_p|c'_p|=k^2/2$ and $\|b_i\|\leq B$, this gives
$\|v_\pi\|\leq Bk^2/2$ and $\|\xi_\pi\|\leq Bk^3\eta^2/6$.  Uniformly over
permutations,
\begin{align}
 \left|\|w_\pi\|^2-\frac{\eta^2}{4}\|v_\pi\|^2\right|
 &\leq \eta\|v_\pi\|\|\xi_\pi\|+\|\xi_\pi\|^2 \notag\\
 &\leq\frac{B^2k^5}{12}\eta^3
 +\frac{B^2k^6}{36}\eta^4
 <\frac{\eta^2}{100}.
 \label{eq:positive-hardness-embedding-error}
\end{align}

If the Partition instance is positive,
Equation~\ref{eq:positive-hardness-pairing-identity} has minimum $4N\Omega^2$.
Otherwise, every signed sum of the even $s_i$ has magnitude at least two, so
the minimum is at least $4N\Omega^2+16$.  The rational threshold
\[
 \theta=\eta^2(N\Omega^2+2)
\]
and Equation~\ref{eq:positive-hardness-embedding-error} strictly separate the
two cases.  All constructed rationals have polynomial bit length.  Given a
permutation, its rational endpoint and squared norm can be evaluated exactly
with polynomially many bit operations, proving membership in NP.

Finally, the opposite request pairs give $\sum_ib_i=0$, so with
$L_R(w)=\tfrac12\|w\|^2$ and $L_{F_i}(w)=\tfrac12\|w+b_i/\eta\|^2$ both the
full-data objective $L_R+\sum_iL_{F_i}$ and the final retain objective $L_R$
have unique minimizer zero, while
$w-\eta\nabla(2L_R(w)-L_{F_i}(w))=(1-\eta)w+b_i$.  Each quadratic is an
identity-design least-squares block, so the hard instance has an explicit
fixed-anchor retain--forget implementation whose update objective has Hessian
$I$, positive response, and step size strictly between zero and one.  The
construction has $d=N+1=k/2+1$ and an instance-dependent $\rho$ approaching
one, and its fixed final-retain anchor is distinct from the prefix-changing
retain loader used by NegGrad+.

This boundary differs from the target-composition hardness of
\citet{kubo2024composition}.  Their strong-hardness construction for increasing
linear functions uses translations $x\mapsto x+a_i$ together with maps whose
slope is larger than one.  It therefore does not directly cover the shared
strict-contraction subclass here.  Conversely, our \textsc{Partition}
reduction is only weakly NP-hard, uses growing dimension, and lets $\rho$
approach one; it isolates realizability by shared positive gradient steps
rather than superseding their broader result.

The construction and the small-step tractable regime are two sides of one
expansion.  A nonzero terminal anchor lets the first-order positional term
dominate the $O(\eta^2)$ interactions, which is what makes small steps
tractable there.  Taking $w_0=z_R=0$ removes that anchor, so a positive
non-overshooting step alone does not imply easy exact optimization.

\subsection{Memory bounds and structural separation}
\label{app:memory-bounds}

\paragraph{Nonlinear common-suffix proof.}
For $u,v\in\mathcal X_s$, composition of the suffix maps gives
\[
 \|U_s(u)-U_s(v)\|\leq\kappa_s\|u-v\|\leq\kappa_sd_s.
\]
Applying the state objective's Lipschitz constant on $U_s(\mathcal X_s)$ gives
the general bound $L_{j,s}\kappa_sd_s$.  For the squared objective the output
difference is at most $L_\Phi\kappa_sd_s$, and writing
$a=\Phi(U_s(u))-z_R$ and $b=\Phi(U_s(v))-z_R$,
\[
 \big|\|a\|^2-\|b\|^2\big|
 =|\langle a+b,a-b\rangle|
 \leq(\|a\|+\|b\|)\|a-b\|
 \leq2R_sL_\Phi\kappa_sd_s.
\]
Maximizing first within each suffix and then over suffixes proves
Proposition~\ref{prop:nonlinear-common-suffix} with $L_{j,s}=2R_sL_\Phi$.
The Lipschitz constants $\nu_{s,t}$ are available for the update rules used here.  For one gradient
step of size $\eta$ on a region with
$\mu I\preceq\nabla^2\ell\preceq\Lambda I$,
$\nu\leq\max\{|1-\eta\mu|,|1-\eta\Lambda|\}$.

\paragraph{Affine suffix truncation and quadratic representation.}
Under the affine assumptions of
Proposition~\ref{prop:finite-memory-ordering}(b), retain only the final
$L$ response terms:
\begin{align}
 \widetilde w_{\pi,L}
 &=M^kw_0+\sum_{r=0}^{L-1}M^r b_{\pi_{k-r}},
 &
 \widetilde J_L(\pi)
 &=\|\widetilde w_{\pi,L}-z_R\|_2^2. \notag
\end{align}
The omitted prefix contribution is
\[
 h_{\pi,L}
 :=w_\pi-\widetilde w_{\pi,L}
 =\sum_{r=L}^{k-1}M^rb_{\pi_{k-r}}.
\]
Using $\|M\|\leq\rho<1$ and $\|b_i\|\leq B$,
\begin{equation}
 \|h_{\pi,L}\|
 \leq
 \delta_L
 :=
 B\sum_{r=L}^{k-1}\rho^r
 =
 B\frac{\rho^L(1-\rho^{k-L})}{1-\rho}.
 \label{eq:squared-finite-memory-definitions}
\end{equation}
Moreover,
\[
 \|\widetilde w_{\pi,L}-z_R\|
 \leq
 \|M^kw_0-z_R\|
 +B\sum_{r=0}^{L-1}\rho^r
 \leq
 \|M^kw_0-z_R\|+\frac{B}{1-\rho}.
\]
Since $w_\pi=\widetilde w_{\pi,L}+h_{\pi,L}$,
\begin{equation}
\begin{aligned}
 |J(\pi)-\widetilde J_L(\pi)|
 &=
 \left|
 2\langle \widetilde w_{\pi,L}-z_R,h_{\pi,L}\rangle
 +\|h_{\pi,L}\|^2
 \right| \\
 &\leq
 2\left(
 \|M^kw_0-z_R\|+\frac{B}{1-\rho}
 \right)\delta_L
 +\delta_L^2
 =:\varepsilon_L .
\end{aligned}
\label{eq:squared-finite-memory-uniform}
\end{equation}
Hence
$\|J-\widetilde J_L\|_\infty\leq\varepsilon_L=O(\rho^L)$,
and the uniform regret bound gives
Proposition~\ref{prop:finite-memory-ordering}(b).

Importantly, truncation removes dependence on the distant prefix without
removing interactions among the retained suffix requests.  Writing
\[
 \widetilde w_{\pi,L}-z_R
 =
 v+\sum_{r<L}d_{\pi_{k-r},r},
\]
and expanding the squared norm gives
\begin{equation}
 \widetilde J_L(\pi)=\|v\|^2
 +\sum_{r=0}^{L-1}
 \left(
 2v^\top d_{\pi_{k-r},r}
 +\|d_{\pi_{k-r},r}\|^2
 \right)
 +2\sum_{0\leq r<s<L}
 d_{\pi_{k-r},r}^\top d_{\pi_{k-s},s}.
 \label{eq:squared-finite-memory-expansion}
\end{equation}
Thus suffix localization suppresses distant-prefix dependence while
preserving arbitrary pairwise interactions within the retained suffix.
\paragraph{Suffix-conditioned assignment.}
Fix an ordered length-$L$ suffix $s$ and write
$\pi=(\sigma,s)$, where $\sigma$ contains the remaining $k-L$ requests.
Conditioning on $s$ fixes both the suffix response and its interaction with
the initial state.  Let
\[
 r_s
 =M^kw_0-z_R+\sum_{j<L}M^jb_{s_{L-j}},
 \qquad
 e_p(i)=M^{k-p}b_i ,
\]
so that
\[
 w_\pi-z_R
 =
 r_s+\sum_{p\le k-L}e_p(\sigma_p).
\]
Expanding the squared norm separates the prefix dependence into unary and
pairwise terms:
\begin{equation}
 J(\pi)=\|r_s\|^2
 +\sum_{p\le k-L}
 \left(
 2r_s^\top e_p(\sigma_p)+\|e_p(\sigma_p)\|^2
 \right)
 +2\sum_{p<q\le k-L}
 e_p(\sigma_p)^\top e_q(\sigma_q).
 \label{eq:suffix-conditioned-split}
\end{equation}
The first two groups form an assignment objective over the prefix, with
suffix-conditioned costs
\[
 C^{(s)}_{i,p}
 =
 2r_s^\top e_p(i)+\|e_p(i)\|^2.
\]
We therefore define
\[
 \widehat J_L(\pi)
 =
 \|r_s\|^2+\sum_p C^{(s)}_{\sigma_p,p},
\]
which keeps the full unary effect of every prefix request and discards only
prefix--prefix interactions.

This conditioning is sharper than truncating the prefix.  Every prefix
position satisfies $k-p\ge L$, so under $\|M\|\le\rho<1$ and
$\|b_i\|\le B$,
\[
 \|e_p(i)\|
 \le
 B\rho^{k-p}
 \le
 B\rho^L.
\]
Hence each discarded interaction contains two responses that have each
passed through at least $L$ further updates.  Using Cauchy--Schwarz and the
geometric series,
\begin{align}
 \left|J(\pi)-\widehat J_L(\pi)\right|
 &=
 2\left|
 \sum_{p<q}
 e_p(\sigma_p)^\top e_q(\sigma_q)
 \right| \\
 &\le
 2B^2
 \sum_{L\le u<t\le k-1}
 \rho^{t+u} \\
 &\le
 B^2
 \left(\sum_{t\ge L}\rho^t\right)^2
 =
 \frac{B^2\rho^{2L}}{(1-\rho)^2}.
\end{align}
Thus conditioning on the suffix improves the affine approximation rate from
$O(\rho^L)$ for prefix truncation to $O(\rho^{2L})$ by retaining the unary
prefix effects and discarding only interactions between already attenuated
prefix responses.

Finally, there are
$k!/(k-L)!\le k^L$ ordered suffixes, and for each one the remaining prefix
requires a single $(k-L)\times(k-L)$ assignment solve.  The resulting search
therefore costs $O(k^{L+3})$ for fixed $L$.

\paragraph{Rate check for truncation and conditioning.}
\label{app:rate-check}
The two exponents are claims about a rate, which the executed study tests at a
single depth.  We check them directly under exactly the shared-affine
assumptions of Proposition~\ref{prop:common-suffix-memory}, on $40$ synthetic
instances at $k=30$ with a five-dimensional state, fitting $\log\varepsilon_L$
against $L\log\rho$ over $L\leq k/3$, where the prefix is still long enough for
the rate to be observable (Table~\ref{tab:suffix-conditioned-rate}).  The
fitted exponents follow the predicted separation across all tested
contractions, and the certificate $B^2\rho^{2L}/(1-\rho)^2$ holds at every
tested length rather than only asymptotically.  The errors are suprema over
$2{,}000$ sampled orders, which lower-bound the supremum over $S_k$.

\begin{table}[H]
\centering
\small
\caption{\textbf{The two affine decay exponents come out as predicted.}
Fitted slopes of $\log\varepsilon_L$ against $L\log\rho$ at $k=30$, over
$40$ instances per contraction.  Proposition~\ref{prop:common-suffix-memory}
predicts $1$ for truncation, $2$ for suffix-conditioned assignment, and hence a
ratio of $2$.}
\label{tab:suffix-conditioned-rate}
\begin{tabular}{@{}rrrr@{}}
\toprule
$\rho$ & Truncation & Conditioning & Ratio \\
\midrule
0.3 & 0.999 & 1.999 & 2.000 \\
0.5 & 1.006 & 2.002 & 1.991 \\
0.7 & 1.007 & 2.011 & 1.998 \\
\bottomrule
\end{tabular}%

\end{table}

The rate comparison above holds at a fixed suffix depth.  Reaching the same
$O(\rho^{2L})$ error by truncating directly to depth $2L$ instead costs
$O(k^{2L})$, so conditioning has the smaller exponent only for $L>3$, with
equality at $L=3$.  The case for conditioning at small $L$ is therefore that it
attains a rate the same depth of truncation does not, not that it is the
cheaper route to that rate.
\paragraph{Queue-length-uniform bounds.}
The absolute errors above are controlled by contraction, but the normalized
structural errors also divide by the decision scale $\Delta_J$.  To separate
these two effects, define the observable last-position signal
\begin{equation}
 \mu_1(k)
 :=
 \mathbb E_{\pi}J(\pi)
 -
 \min_i \mathbb E\!\left[J(\pi)\mid \pi_k=i\right].
 \label{eq:decision-scale}
\end{equation}
It measures how much the expected objective can improve by choosing only the
final request.  Since every conditional mean is at least the global minimum,
\[
 \Delta_J
 =
 \mathbb E_\pi J(\pi)-\min_\pi J(\pi)
 \geq
 \mu_1(k).
\]
Thus $\mu_1(k)$ is an observable lower bound on the normalization.

\paragraph{Suffix width.}
If two orders $\pi$ and $\pi'$ share the same ordered length-$L$ suffix,
then $\widetilde J_L(\pi)=\widetilde J_L(\pi')$.  Equation~\ref{eq:squared-finite-memory-uniform}
therefore gives
\[
 |J(\pi)-J(\pi')|
 \le
 |J(\pi)-\widetilde J_L(\pi)|
 +
 |J(\pi')-\widetilde J_L(\pi')|
 \le 2\varepsilon_L .
\]
Taking the maximum over orders with a common suffix yields
\begin{equation}
 W_L\leq 2\varepsilon_L.
 \label{eq:width-two-epsilon}
\end{equation}

\begin{corollary}[Queue-length-uniform structural bounds]
\label{cor:queue-uniform}
For an ordered length-$L$ suffix $s$, let $\mathcal C_L^{\rm cond}$ contain
functions of the form
$f(\sigma s)=c_s+\sum_{p\leq k-L}C^{(s)}_{\sigma_p,p}$, and define
\[
 D_{\rm cond}(L)
 :=\frac{2}{\Delta_J}
 \inf_{f\in\mathcal C_L^{\rm cond}}\|J-f\|_\infty.
\]
Under Proposition~\ref{prop:common-suffix-memory}(b), there exists a constant
$c_\rho$, depending on $\rho$, $B$, and
$R_0=\sup_k\|M^kw_0-z_R\|$, but not on $k$ or $L$, such that
\begin{equation}
 D_{S,L}
 \leq
 \frac{2c_\rho\rho^L}{\mu_1(k)},
 \qquad
 D_{\rm cond}(L)
 \leq
 \frac{2B^2\rho^{2L}}
 {(1-\rho)^2\mu_1(k)}.
 \label{eq:queue-uniform}
\end{equation}
The second bound applies to the suffix-conditioned family of
Proposition~\ref{prop:common-suffix-memory}(c).  At $L=0$, this family reduces
to assignment, giving
\[
 D_{\mathcal A}
 \leq
 \frac{2B^2}
 {(1-\rho)^2\mu_1(k)}.
\]
If, along a sequence of instances, the response constants remain uniformly
bounded and the decision scale stays non-degenerate,
$\inf_k\mu_1(k)=m>0$, then any fixed normalized tolerance $\tau$ is achieved
with
\[
 L
 \geq
 \frac{\log(2c_\rho/(m\tau))}{\log(1/\rho)}
 =
 O(\log(1/\tau)),
\]
independently of $k$.
\end{corollary}

The proof is immediate from the previous bounds.  Equation~\ref{eq:width-two-epsilon}
gives $W_L\leq2\varepsilon_L$.  Equation~\ref{eq:squared-finite-memory-definitions}
gives $\delta_L\leq B\rho^L/(1-\rho)$, and $\|M\|\leq\rho<1$ keeps
$R_0=\sup_k\|M^kw_0-z_R\|$ finite, so
Equation~\ref{eq:squared-finite-memory-uniform} collapses to
$\varepsilon_L\leq c_\rho\rho^L$ with $c_\rho$ free of $k$ and $L$.  Dividing by
$\Delta_J\geq\mu_1(k)$ yields the first inequality.
Likewise, dividing Equation~\ref{eq:suffix-conditioned-error} by the same
lower bound gives the second.

The non-degeneracy condition is essential.  Contraction controls the numerator,
whereas $\mu_1(k)$ controls whether the normalized decision scale remains
visible as the queue grows.  If every request behaves identically in the final
position, then $\mu_1(k)=0$ and the bound becomes uninformative.  This does not
imply that ordering itself is irrelevant, only that last-position variation no
longer lower-bounds $\Delta_J$.

The relevant scaling variables are therefore the contraction and the decision
scale, rather than queue length by itself.  Appending requests while keeping a
fixed per-request contraction leaves the bound uniform in $k$.  If the update
instead becomes increasingly close to the identity as $k$ grows, the generic
argument no longer supplies a queue-length-independent depth, and the factor
$(1-\rho_k)^{-2}$ also worsens.  This deterioration of the upper bound is not
an instance-wise lower bound on the suffix depth required.

The same separation extends beyond affine dynamics.  Under
Proposition~\ref{prop:common-suffix-memory}(a),
\[
 D_{S,L}
 \leq
 \frac{
 2L_\Phi\max_s(R_s\kappa_sd_s)
 }{
 \mu_1(k)
 }.
\]
Hence a policy family with $R_sd_s\leq C$ and $\kappa_s\leq q^L$ uniformly in
$k$ inherits the same queue-length-uniform conclusion.  We do not certify such
uniform constants for the executed full-network policy.  The larger-queue study
therefore measures these quantities empirically rather than invoking this
corollary.
\paragraph{Strict separation between memory and additivity.}
Neither suffix localization nor position additivity implies the other, even
under shared affine dynamics.  We show the two directions by explicit
constructions.

\emph{Short memory does not imply additivity.}
Let $w_0=z_R=0$ and
\begin{equation}
 M=\begin{pmatrix}0&1/2\\0&0\end{pmatrix},
 \qquad
 b_i=(x_i,x_i)^\top,
 \qquad
 (x_1,x_2,x_3,x_4)=(-2,1,-1,2).
 \label{eq:finite-memory-separation-instance}
\end{equation}
Since $M^2=0$, only the final two requests affect the terminal state:
\[
 w_\pi=Mb_{\pi_3}+b_{\pi_4}.
\]
Writing $a=x_{\pi_3}$ and $b=x_{\pi_4}$ gives
\begin{equation}
 J(\pi)=2b^2+ab+\frac14a^2.
 \label{eq:finite-memory-separation-objective}
\end{equation}
Thus $J$ is constant over all prefixes sharing the same ordered two-request
suffix, and hence $W_2=0$.  The remaining suffix dependence is nevertheless
not additive.  Consider
\[
 \pi^{(1)}=(2,4,3,1),\quad
 \pi^{(2)}=(1,3,4,2),\quad
 \pi^{(3)}=(2,3,4,1),\quad
 \pi^{(4)}=(1,4,3,2).
\]
Any assignment objective
$f(\pi)=c+\sum_p C_{\pi_p,p}$ satisfies
\begin{equation}
 f(\pi^{(1)})+f(\pi^{(2)})
 -f(\pi^{(3)})-f(\pi^{(4)})=0,
 \label{eq:assignment-four-order-identity}
\end{equation}
because every request--position term cancels in this signed contrast.
For the true objective, Equation~\ref{eq:finite-memory-separation-objective}
gives the corresponding values
$41/4,5,5,$ and $5/4$, whose signed contrast is $9$.
Therefore, if
$\varepsilon=\max_\pi |J(\pi)-f(\pi)|$, the triangle inequality gives
$9\le 4\varepsilon$, and hence
\begin{equation}
 \inf_{f\in\mathcal A}\max_\pi|J(\pi)-f(\pi)|
 \ge \frac94.
 \label{eq:finite-memory-assignment-separation}
\end{equation}
Direct enumeration gives
\[
 \mathbb E_\pi J(\pi)=\frac{115}{24},
 \qquad
 \min_\pi J(\pi)=1,
 \qquad
 \Delta_J=\frac{91}{24},
\]
so
\[
 D_{\mathcal A}
 \ge \frac{108}{91}>1.
\]
Thus exact two-request suffix localization can coexist with a substantial
failure of position additivity.

\emph{Additivity does not imply short memory.}
For the reverse direction, take the one-dimensional instance
\[
 M=\frac45,\qquad
 w_0=0,\qquad
 z_R=1,\qquad
 b_1=1,\qquad
 b_2=b_3=b_4=0.
\]
If request $1$ occupies position $p$, then
\[
 J(\pi)
 =
 \left[
 \left(\frac45\right)^{4-p}-1
 \right]^2.
\]
The objective therefore depends only on the request--position assignment, so
$D_{\mathcal A}=0$.  Its four position values, in units of $1/15625$, are
$3721,2025,625,$ and $0$, which gives
\[
 \Delta_J=\frac{6371}{62500}.
\]
However, fixing an ordered suffix of length two does not determine the
objective: when request $1$ lies outside the suffix, it may still occupy
prefix position $1$ or $2$.  Consequently,
\[
 W_2
 =
 \frac{3721-2025}{15625}
 =
 \frac{1696}{15625},
 \qquad
 \frac{W_2}{\Delta_J}
 =
 \frac{6784}{6371}>1.
\]
Hence exact position additivity can coexist with substantial residual
dependence on the prefix.
\subsection{From assignment to scalar sorting}
\label{app:auxiliary-projections}

\paragraph{Scope of the slice projection.}
Proposition~\ref{prop:slice-additive-certificate} certifies position
additivity through one specific route, an independent additive projection of
each request-pair slice.  A small residual is therefore sufficient for
approximate position additivity but not necessary.  A coefficient tensor can be
linearizable by a construction this projection does not find, so a large
residual leaves the case uncertified rather than settled.

\paragraph{How tight is the sorting certificate?}
\label{app:certificate-tightness}
Equation~\ref{eq:q-regret} is a sufficient bound, so a large value leaves a
case uncertified rather than unstructured.  Table~\ref{tab:sorting-certificate}
compares best-family error, the specific sorting surrogate, its certificate,
and realized decision regret on the complete $k=4$ surfaces.  The certificate
is conservative where the family fits.  The worst softmax cases nevertheless
lose more than the random-order gap, whereas the reported SalUn cases do not;
median performance therefore does not make sorting uniformly safe.

Double centering does not change the decision, only what the bound charges for
it.  Charging the position-only effect to the residual inflates the certificate
by a median of $81\%$ across the $95$ complete surfaces the audit refits, with
a range of $5$ to $99\%$, while leaving every selected order, every realized
regret, and every centered score unchanged.

\begin{table}[H]
\centering
\small
\setlength{\tabcolsep}{4pt}
\caption{\textbf{The sorting certificate is conservative where the family
fits.}  Entries are medians with observed $[\min,\max]$ across the complete
$k=4$ surfaces of each setting, all normalized by the random-order gap.
$D_{\mathcal A}$ is the best assignment family's uniform error, the sorting
error is that of the surrogate $sq^\top$ itself, the certificate is the
right-hand side of Equation~\ref{eq:q-regret}, and the realized regret is what
the sorted order loses on the complete table.}
\label{tab:sorting-certificate}
\begin{tabular}{@{}lrrrrr@{}}
\toprule
Setting & Inst. & $D_{\mathcal A}$ & Sorting error & Certificate & Realized regret \\
\midrule
Frozen ridge, vision & 15 & \shortstack[r]{$0.010$\\$[0.001,0.066]$} & \shortstack[r]{$0.059$\\$[0.015,0.266]$} & \shortstack[r]{$0.080$\\$[0.021,0.375]$} & \shortstack[r]{$0.000$\\$[0.000,0.000]$} \\
Frozen ridge, AG News & 15 & \shortstack[r]{$0.018$\\$[0.002,0.068]$} & \shortstack[r]{$0.062$\\$[0.017,0.319]$} & \shortstack[r]{$0.082$\\$[0.023,0.446]$} & \shortstack[r]{$0.000$\\$[0.000,0.092]$} \\
Frozen ridge, 20 Newsgroups & 15 & \shortstack[r]{$0.022$\\$[0.002,0.126]$} & \shortstack[r]{$0.074$\\$[0.015,0.321]$} & \shortstack[r]{$0.116$\\$[0.022,0.473]$} & \shortstack[r]{$0.000$\\$[0.000,0.000]$} \\
Direct softmax head & 15 & \shortstack[r]{$0.035$\\$[0.005,0.634]$} & \shortstack[r]{$0.081$\\$[0.017,2.616]$} & \shortstack[r]{$0.130$\\$[0.021,4.005]$} & \shortstack[r]{$0.000$\\$[0.000,1.248]$} \\
Factorized softmax head & 15 & \shortstack[r]{$0.065$\\$[0.010,0.686]$} & \shortstack[r]{$0.584$\\$[0.117,2.523]$} & \shortstack[r]{$0.872$\\$[0.184,4.059]$} & \shortstack[r]{$0.009$\\$[0.000,1.789]$} \\
NegGrad+ & 10 & \shortstack[r]{$0.236$\\$[0.119,0.391]$} & \shortstack[r]{$1.867$\\$[1.189,2.998]$} & \shortstack[r]{$2.672$\\$[1.807,4.966]$} & \shortstack[r]{$0.122$\\$[0.000,0.309]$} \\
SalUn & 10 & \shortstack[r]{$0.673$\\$[0.215,1.030]$} & \shortstack[r]{$1.734$\\$[1.225,2.686]$} & \shortstack[r]{$2.287$\\$[1.655,4.297]$} & \shortstack[r]{$0.322$\\$[0.000,0.775]$} \\
\bottomrule
\end{tabular}%

\end{table}

\paragraph{Linear-position projection proofs.}

For $p=1,\ldots,k$, let $q_p=p-(k+1)/2$, so $\|q\|_2^2=k(k^2-1)/12$, and let
$H=I-\tfrac1k\mathbf 1\mathbf 1^\top$.  Writing $a$ for the row means of
$C$, $b$ for its column means and $m$ for its grand mean,
\[
 C=HCH+a\mathbf 1^\top+\mathbf 1b^\top-m\mathbf 1\mathbf 1^\top,
\]
and for any permutation
$\Phi_{a\mathbf 1^\top}(\pi)=\sum_ia_i$,
$\Phi_{\mathbf 1b^\top}(\pi)=\sum_pb_p$ are constants, because each request
and each position is used exactly once.  Comparing permutations under $C$ is
therefore the same as comparing them under $\bar C=HCH$.  Write
$\bar C=P+\bar\Xi_q$ with $P=s q^\top$ and
$\bar\Xi_q\mathbf 1=\bar\Xi_q q=0$, which also leaves $\bar\Xi_q$
column-centered since $\bar C$ is.  For any permutation $\pi$,
\[
 \Phi_P(\pi)=\sum_p s_{\pi_p}q_p.
\]
Because $q_p$ is strictly increasing, the rearrangement inequality shows that
sorting $s_i$ nonincreasingly minimizes $\Phi_P$.  This proves exact scalar
collapse when $\bar\Xi_q=0$.

The centering never loosens the certificate.  If $C=a\mathbf 1^\top+sq^\top
+\Xi_q$ is the one-sided projection, then applying $H$ on both sides gives
$\bar C=(Hs)q^\top+H\Xi_qH$, since $Hq=q$ and $\mathbf 1^\top H=0$.  As
$\bar\Xi_q$ is the residual of the best such fit and $H$ is an orthogonal
projection,
$\|\bar\Xi_q\|_F\leq\|H\Xi_qH\|_F\leq\|\Xi_q\|_F$.  The two
projections also induce the same ordering, because they differ in $s$ only by
a constant shift, which the rearrangement inequality ignores.

For the approximate result, let $\pi_C$ minimize $\Phi_{\bar C}$ and $\pi_s$
minimize $\Phi_P$.  Then
\[
 \Phi_{\bar C}(\pi_s)-\Phi_{\bar C}(\pi_C)
 \leq \Phi_{\bar\Xi_q}(\pi_s)-\Phi_{\bar\Xi_q}(\pi_C)
 \leq \operatorname{range}_\pi \Phi_{\bar\Xi_q}(\pi).
\]
If $X_\pi$ is the permutation matrix selecting the entries summed by $\pi$,
then
\[
 \Phi_{\bar\Xi_q}(\pi)-\Phi_{\bar\Xi_q}(\sigma)
 =\langle \bar\Xi_q,X_\pi-X_\sigma\rangle
 \leq \|\bar\Xi_q\|_F\|X_\pi-X_\sigma\|_F
 \leq \sqrt{2k}\,\|\bar\Xi_q\|_F.
\]
The last inequality uses that two permutation matrices differ in at most
$2k$ unit entries.  This bounds the regret of $\pi_s$ against the assignment
objective $\Phi_C$.  Adding the two uniform errors between $J$ and $\Phi_C$, one
at the selected order and one at the true optimum, gives
Equation~\ref{eq:q-regret} with $\varepsilon_{\mathcal A}$ in place of the
surrogate mismatch.  Chaining part (a) supplies
$\varepsilon_{\mathcal A}\leq\varepsilon_{\rm surr}
+\sqrt{\binom{k}{2}}\,\|\Xi\|_F$, since the projected assignment differs
from the shared quadratic surrogate by the interaction residual and the
surrogate differs from $J$ by $\varepsilon_{\rm surr}$.  The
alternative entrywise bound is
$\operatorname{range}_\pi \Phi_{\bar\Xi_q}(\pi)
\leq2k\|\bar\Xi_q\|_\infty$.

\section{Experimental protocols}
\label{app:protocols}

This section defines how each experimental setting is executed and which
executions count as valid.  A \emph{setting} fixes a policy, model, and data
configuration, while an \emph{instance} additionally fixes the requests,
checkpoint, reference, and request-bound randomness.  Permutations are
deletion orders within an instance.  Complete permutation surfaces support
uniform statements over $S_k$, whereas identification and held-out splits
support only statements on their declared samples.  Frozen-head objectives
are deterministic in double precision, and fixed-randomness repeats reproduce
the full-network terminal objectives exactly.  Results are reported in
Appendices~\ref{app:structural-evidence} and~\ref{app:identification}.

Table~\ref{tab:fullnet-instance-map} records which full-network analyses reuse
an instance.  Seed labels are local identifiers: the same number at different
queue lengths does not denote the same request set.

\begin{table}[H]
\centering
\small
\setlength{\tabcolsep}{3.5pt}
\caption{\textbf{Full-network instance correspondence.}  ``Main'' means the
complete $k=4$ cohort used for structural summaries.}
\label{tab:fullnet-instance-map}
\begin{tabular}{@{}lll@{}}
\toprule
Analysis & Instances & Relation to main $k=4$ cohort \\
\midrule
NegGrad+ budget development & seeds 0--2, $k=4$ & Same requests; alternate budgets \\
NegGrad+ main & seeds 0--9, $k=4$ & Complete 24-order surfaces \\
SalUn pilot/replication & seeds 0--2 / 3--9, $k=4$ & Matched request partitions by seed \\
Composition & seeds 0--2, $k=5$ & New five-request queues \\
Held-out identification & seeds 0--2, $k=8$ & New eight-request sampled designs \\
Second tape / reference-free & seeds 0--2, $k=4$ & Same requests/reference; separate $10{,}000$-audit replays \\
Pooled policy control & seeds 0--9, $k=4$ & Reuse NegGrad+ main instances \\
\bottomrule
\end{tabular}%

\end{table}

\paragraph{Sampled versus uniform claims.}
When $S_k$ is too large to enumerate, the largest residual observed on a
sampled evaluation set is only a sample-level statistic.  It does not certify
the maximum over all permutations.  We therefore restrict sampled errors to
their declared evaluation sets.  Uniform quantities such as
$D_{\mathcal A}$ and $D_{S,L}$ inherit the uniform regret guarantee only when
their underlying error is controlled uniformly.  At large $k$, such uniform
control comes from Section~\ref{sec:convex-theory}, not from treating a sampled
maximum as exhaustive.

\paragraph{Information used by each construction.}
\label{app:information-conditions}
The structural constructions use different information regimes.  A
\emph{dynamics-based} construction follows directly from the update equations.
A \emph{reference-assisted} construction uses terminal objectives together
with a retraining reference to fit or evaluate a structural family.  A
\emph{probe-based} construction instead uses the base model, request data,
retained-data derivatives, and declared singleton probes.  Constructions fixed
on development executions before a new queue are prospective on that queue.
Complete-table fits are retrospective.

We separate reference generation, identification executions, offline fitting
and solving, and candidate execution.  One terminal-order query executes the
full $k$-request sequence and evaluates its terminal objective.  Identification
executions are disjoint from held-out or newly executed evaluation orders.
Table~\ref{tab:information-availability} reports counts, not FLOPs or
wall-clock time; those depend on batch composition, forward/backward work, and
evaluation and were not benchmarked here.

\begin{table}[H]
\centering
\small
\setlength{\tabcolsep}{4pt}
\caption{\textbf{Information and execution costs by construction.}
``Seq.'' counts terminal-order executions.  Parentheses give optimizer updates
when they follow directly from the protocol.  Reference-assisted workflows
include reference generation and are not claimed to be cheaper than
retraining.}
\label{tab:information-availability}
\begin{tabular}{@{}lllll@{}}
\toprule
Construction & $T_{\rm reference}$ & $T_{\rm identification}$
 & $T_{\rm fit/solve}$ & $T_{\rm candidate}$ \\
\midrule
Ridge sorting & Exact retain solve & $k$ seq. & Least squares, sort & $1$ seq. \\
Ridge assignment & Exact retain solve & $(k-1)^2+1$ seq. & Minimax, Hungarian & $1$ seq. \\
NegGrad+ $k=8$ & Two retain-only runs & $112$ seq. ($10{,}752$) & Minimax, solve & $1$ seq. ($96$) \\
Head reconstruction & Exact retain solve & $0$ seq.; $k$ probes & Construct, solve & $1$ seq. \\
\bottomrule
\end{tabular}
\end{table}

\subsection{Frozen heads}

Ridge provides an exact retained reference for measuring endpoint prediction
error and ordering regret.  CIFAR-10 uses normalized penultimate features from
a public ImageNet-pretrained ResNet-18 and a two-output ridge head with bias and
$\lambda=0.01$.  AG News and 20 Newsgroups use mean-pooled frozen DistilBERT
embeddings, a shared data-independent 256-dimensional projection, and
per-example normalization.  The protocol fixes request size,
train/validation splits, and the retained reference before selection.

\subsection{Full-network NegGrad+}

\paragraph{Terminal comparison map.}
For each main-cohort instance, the first 2,000 examples of the fixed CIFAR-100
test split form the audit set; the remainder is used only for reported
accuracies.  The classifier predicts the 20 CIFAR-100 superclasses.  If
$z_w(x)\in\mathbb R^{20}$ denotes its logits, the comparison map subtracts the
mean across the 20 coordinates separately for each example.  The terminal
objective is
\[
 j(w)=\frac{1}{|\mathcal A|}\sum_{x\in\mathcal A}
 \left\|z_w(x)-\overline z_w(x)\mathbf 1-
 \bigl(z_R(x)-\overline z_R(x)\mathbf 1\bigr)\right\|_2^2.
\]
Here $z_R$ is the pointwise mean of two models trained from the declared
initialization on the final retained training set.  The audit examples and
reference are fixed across every order of an instance.  Only training data are
used for request updates; the test audit set is read at terminal evaluation.
The second-tape and reference-free controls reuse the seed-indexed requests and
reference files but come from separate replays evaluated on all 10,000 test
examples.  They are therefore distinct policy--objective instances from the
main 2,000-example cohort, even when their request labels match.

\paragraph{Execution.}
Each request uses 12 NegGrad+ steps with learning rate $3\times10^{-3}$ and
minibatch size 256.  The loss weights retained and forget cross-entropy by
$0.95$ and $-0.05$, respectively.  Each request starts a fresh SGD optimizer
with momentum $0.9$ and weight decay $5\times10^{-4}$.  The updated
model weights
and batch-normalization state pass to the next request.  Retain batches use the
run's CIFAR augmentation and forget batches do not.

\paragraph{Operating point and validity.}
\label{app:update-strength}
The 12-step operating point was fixed by the preceding pilot calibration using
retained utility and order-mean progress toward the retraining reference; no
ordering statistic or deleted-group screen selected it.  The four-budget
$\{3,12,24,48\}$ sweep was run afterward with model, requests, reference, and
request-bound randomness held fixed.  Under the three screens reported here,
12 steps is the only tested budget that passes on all three development
instances.  The deleted-group screen is therefore a retrospective audit, not a
backdated selection rule.  Per-instance values for the retain-weight sweep at
fixed step size are in the repository.

\paragraph{Structure across the budget sweep.}
\label{app:budget-structure}
The same four budgets also carry structural coordinates, which the selection
did not look at (Table~\ref{tab:update-strength-sweep}).  Nominal update count
does not order suffix attenuation monotonically, and assignment error has no
corresponding trend.  The effective suffix depth changes across the sweep, but
all alternative budgets fail at least one deletion-quality screen.  We
therefore treat this as operating-region sensitivity and ground the causal
mechanism claim on the damped-Newton intervention of
Appendix~\ref{app:newton-sweep}, where contraction is fixed before execution.

\begin{table}[H]
\centering
\small
\caption{\textbf{Structural coordinates across the update-strength sweep.}
Entries are medians with observed $[\min,\max]$ over the same three
development instances, on complete $k=4$ surfaces.  ``Valid seeds'' counts the
instances passing all three reported screens.  History retention is the measured
matched-history attenuation, and $D_{S,1}$ and $D_{S,2}$ are the suffix errors
at one and two positions.}
\label{tab:update-strength-sweep}
\begin{tabular}{@{}rrrrrr@{}}
\toprule
Steps/request & Valid seeds & History retention & $D_{S,1}$ & $D_{S,2}$ & $D_{\mathcal A}$ \\
\midrule
3 & 0/3 & \shortstack[r]{$0.318$\\$[0.312,0.321]$} & \shortstack[r]{$1.362$\\$[0.843,1.399]$} & \shortstack[r]{$0.515$\\$[0.341,0.531]$} & \shortstack[r]{$0.443$\\$[0.260,0.512]$} \\
12 & 3/3 & \shortstack[r]{$0.065$\\$[0.059,0.066]$} & \shortstack[r]{$0.598$\\$[0.573,0.883]$} & \shortstack[r]{$0.221$\\$[0.204,0.286]$} & \shortstack[r]{$0.302$\\$[0.265,0.441]$} \\
24 & 2/3 & \shortstack[r]{$0.118$\\$[0.108,0.127]$} & \shortstack[r]{$0.394$\\$[0.363,0.417]$} & \shortstack[r]{$0.149$\\$[0.117,0.202]$} & \shortstack[r]{$0.221$\\$[0.166,0.229]$} \\
48 & 0/3 & \shortstack[r]{$0.278$\\$[0.225,0.312]$} & \shortstack[r]{$0.993$\\$[0.840,1.167]$} & \shortstack[r]{$0.673$\\$[0.497,0.892]$} & \shortstack[r]{$0.256$\\$[0.198,0.378]$} \\
\bottomrule
\end{tabular}%

\end{table}

At step $t$ the retain loader is the complement of every request deleted
through that step, $R_t=\mathcal D\setminus\bigcup_{s\leq t}F_{\pi_s}$, so
deleted samples are never reintroduced as retain data.  All ten reported
instances satisfy the retained-utility criterion, worst-order retain accuracy
at least $\min(A_{\rm base},A_{\rm retrain})-0.05$, and the reference-progress
criterion, order-mean terminal output closer to the retraining reference than
the base output.  The deleted-group accuracy screen is reported as an
additional retrospective check.  All ten pass, with a smallest decrease of
$0.120$, so the measured rebound reflects recovery under later updates.

\subsection{Full-network SalUn}
\label{app:salun-protocol}

\paragraph{Execution.}
SalUn is executed from the authors' implementation.  Each request uses 12
steps with learning rate $10^{-4}$ and minibatch size 256, a fresh SGD
optimizer per request with momentum $0.9$ and weight decay $5\times10^{-4}$,
retain-only augmentation, and batch-normalization in training mode.  The
updated weights and batch-normalization state pass to the next request, as for
NegGrad+.  At every request, the mask is recomputed from the current model by
accumulating absolute gradients of the negated forget loss and globally
selecting the top $50\%$ of parameter coordinates.  The authors' random-label
update is then applied only on those coordinates.  Its NumPy and PyTorch draws
are seeded from the request-bound generator; forget examples are unaugmented,
and a retained sample fills the fixed $12\times256$-example pass.

\paragraph{Operating point and validity.}
SalUn uses the 12-step budget fixed previously by the full-network protocol
rather than a SalUn-specific sweep.  At this operating point all ten instances
satisfy the same retained-utility, reference-progress and deleted-group
checks, with worst-order retain accuracy $0.847$--$0.898$, reference-distance
ratios of $0.631$--$0.753$, and deleted-group accuracy falling by
$0.114$--$0.175$ against the screen's $0.05$.  No structural
diagnostic enters any of these checks.

\section{Mechanisms and structural diagnostics}
\label{app:structural-evidence}

This appendix carries the evidence for where the observed reductions come from
and the controls that separate them from family flexibility.  The ledgers
below record the diagnostics summarized by Figure~\ref{fig:structural-axes}.

Table~\ref{tab:structural-regimes-complete} compares the structural regimes on
surfaces enumerated in full.  The settings do not share a single sufficient
family: position additivity dominates the frozen-head surfaces, while suffix
localization distinguishes the complete full-network surfaces from the ridge
extension.  Objectives and normalizations are setting-specific, so entries
compare instances within a setting rather than across settings.

For complete permutation surfaces, $D_{\mathcal A}$ is the exact best
assignment-family error
over all $k!$ orders and $D_S$ the exact suffix family error.  The two
sampled designs carry a different quantity and are reported with the
identification results in Appendix~\ref{app:identification}.

\begin{table}[H]
\centering
\footnotesize
\setlength{\tabcolsep}{2.5pt}
\caption{\textbf{Structural regimes on the completely enumerated surfaces.}
Entries are medians across instances with observed $[\min,\max]$.  \emph{Gap}
is $100\Delta_J/\bar J$, every error is exact over all $k!$ orders, and the
last two columns count the instances whose assignment and suffix errors fall
below one.  All suffix fits condition on the final two requests.}
\label{tab:structural-regimes-complete}
\begin{tabular}{@{}lrrrrrcc@{}}
\toprule
Setting & Req. & Inst. & Gap (\%) & $D_{\mathcal A}$ & $D_S$ & A pass & S pass \\
\midrule
Ridge head & 4 & 45 & \shortstack[r]{$0.878$\\$[0.063,11.659]$} & \shortstack[r]{$0.018$\\$[0.001,0.126]$} & \shortstack[r]{$0.562$\\$[0.427,0.658]$} & 45/45 & 45/45 \\
Direct head & 4 & 15 & \shortstack[r]{$0.076$\\$[0.003,1.055]$} & \shortstack[r]{$0.035$\\$[0.005,0.634]$} & \shortstack[r]{$0.589$\\$[0.492,1.700]$} & 15/15 & 13/15 \\
Factorized head & 4 & 15 & \shortstack[r]{$0.811$\\$[0.073,1.446]$} & \shortstack[r]{$0.065$\\$[0.010,0.686]$} & \shortstack[r]{$0.547$\\$[0.383,1.303]$} & 15/15 & 14/15 \\
Ridge head & 8 & 3 & \shortstack[r]{$0.838$\\$[0.824,0.996]$} & \shortstack[r]{$0.017$\\$[0.015,0.021]$} & \shortstack[r]{$1.290$\\$[1.287,1.318]$} & 3/3 & 0/3 \\
NegGrad+ & 4 & 10 & \shortstack[r]{$9.285$\\$[3.407,15.201]$} & \shortstack[r]{$0.236$\\$[0.119,0.391]$} & \shortstack[r]{$0.148$\\$[0.055,0.477]$} & 10/10 & 10/10 \\
NegGrad+ & 5 & 3 & \shortstack[r]{$9.647$\\$[3.660,13.805]$} & \shortstack[r]{$0.186$\\$[0.150,0.771]$} & \shortstack[r]{$0.238$\\$[0.227,0.582]$} & 3/3 & 3/3 \\
SalUn & 4 & 10 & \shortstack[r]{$8.488$\\$[5.831,13.501]$} & \shortstack[r]{$0.673$\\$[0.215,1.030]$} & \shortstack[r]{$0.260$\\$[0.185,0.510]$} & 9/10 & 10/10 \\
\bottomrule
\end{tabular}%

\end{table}

\subsection{Position additivity from update dynamics}
\label{app:position-mechanism}

\paragraph{Cumulative-response construction.}
Let $u_i$ be the singleton displacement and set
$M=(I-\eta H_R)^T$ for a budget of $T$ steps.  A request deleted at position $p$ remains absent in every
later stage, giving the shared-curvature response
\begin{equation}
 e_{i,p}=\sum_{t=0}^{k-p}M^t u_i,
 \qquad p=1,\ldots,k.
 \label{eq:persistent-response}
\end{equation}
For $1\leq p\leq k-2$, direct subtraction yields
\begin{equation}
 e_{i,p+2}-2e_{i,p+1}+e_{i,p}=M^{k-p-1}(M-I)u_i.
 \label{eq:cumulative-curvature}
\end{equation}
The telescoping identity for $M$ implies
\[
 \|M-I\|\leq T\eta\|H_R\|
 \max\{1,\|I-\eta H_R\|\}^{T-1}.
\]
If the singleton gradient norm is at most $G_i$ along its $T$ steps, then
$\|u_i\|\leq T\eta G_i$ and
\begin{equation}
 \|e_{i,p+2}-2e_{i,p+1}+e_{i,p}\|
 \leq T^2\eta^2\|M\|^{k-p-1}\|H_R\|G_i
 \max\{1,\|I-\eta H_R\|\}^{T-1}.
 \label{eq:cumulative-curvature-bound}
\end{equation}
Thus the cumulative response is affine in position up to second order in the
step size for fixed $T$ and $k$.

\paragraph{From the retained recurrence to the terminal objective.}
The executed ridge policy minimizes a mean retained loss, so its normalization
changes with the stage.  Let the full training set have size $N$, let every
request have size $m$, and let $n_t=N-tm$ after stage $t$.  If $r_i$ is the sum
of the unregularized per-example gradients in request $i$, expansion of the
actual $T$-step recurrence at the full-data optimum gives, uniformly over the
finite set of orders,
\begin{equation}
 w_\pi(\eta)=w_0+\eta c+T\eta\sum_{p=1}^k h_p r_{\pi_p}+O(\eta^2),
 \qquad h_p=\sum_{t=p}^k\frac{1}{n_t},
 \label{eq:ridge-first-order-profile}
\end{equation}
where $c$ is independent of $\pi$.  Expanding the squared-distance objective
around $w_0$ therefore gives a scalar request--position objective with profile
$h$ plus an $O(\eta^2)$ remainder.  The exact profile is not fixed-$q$ because
$n_t$ changes.  For fixed $k$,
$h_p=(k-p+1)/N+O(m/N^2)$.  Consequently, projection onto fixed-$q$ has both a
small first-order normalization term and a second-order dynamics term.  If the
first-order order variation is nonzero, $\Delta_J=\Theta(\eta)$, while $D_{\rm sort}$ need not
scale quadratically.  This derivation is specific to the matched ridge
recurrence and does not convert the finite-sample sweep into an asymptotic
certificate.

\paragraph{Matched step-size intervention.}
\label{app:ridge-eta-sweep}
We fixed $T=4$ gradient steps per request, the five held-out CIFAR-10
request partitions, the base ridge
head, and every exact retained-data reference, and varied only the learning
rate over $0.5\eta_0$, $\eta_0$, $1.5\eta_0$, and $2\eta_0$, where
$\eta_0=0.526$.  All 20 matched configurations contained 65 independently replayed prefixes
and 24 finite terminal objective values.  The largest product of learning rate and
retained Hessian eigenvalue was $1.801<2$.  The recurrence-specific residual
has fixed-effect log--log slope $1.971$.  Its near-quadratic growth is
consistent with the positional-curvature calculation in Equation~
\ref{eq:cumulative-curvature-bound} and the second-order remainder in
Equation~\ref{eq:ridge-first-order-profile}.

\begin{figure}[t]
  \centering
  \includegraphics[width=0.72\linewidth]{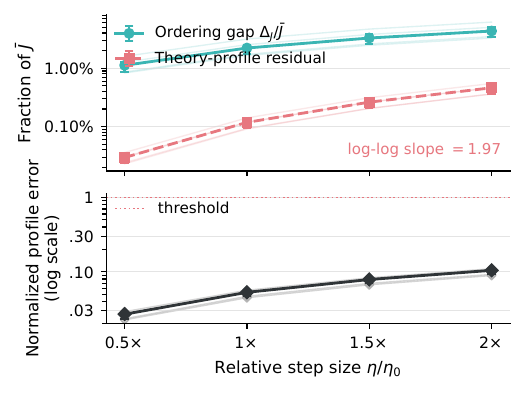}
  \caption{\textbf{Positional-profile error grows at the second-order rate
  predicted by the update expansion.}  The upper panel shows absolute order gap
  and profile residual against learning rate, and the lower panel the same
  residual normalized by the gap.}
  \label{fig:ridge-eta-sweep-full}
\end{figure}

Figure~\ref{fig:ridge-eta-sweep-full} shows every configuration of that sweep, in
absolute terms and normalized by each configuration's gap.  Here $h$ is the
recurrence-specific profile in
Equation~\ref{eq:ridge-first-order-profile}.  Its residual is smaller than the
fixed linear-position profile in all 20 matched partition--step-size
configurations.

\paragraph{Shared-response quadratic closure.}
\label{app:quadratic-closure}
We reused the singleton displacements and shared response from the pre-outcome
analysis, and used the exact retained reference only for this retrospective
comparison.  The linear surrogate scores the predicted displacement by its
first-order inner product with $w_0-w_R$.  The quadratic surrogate instead
evaluates the exact squared norm of that same predicted endpoint, thereby
adding precisely the interactions in Equation~\ref{eq:shared-quadratic}.
Nothing about the response model, requests, or reference changed between the
two columns.
This comparison tests the terminal-objective consequence of the derived shared
response, but does
not assert that the actual ridge
execution has one fixed response matrix.  Its retained Hessian changes with the
deleted request set.  Table~\ref{tab:quadratic-closure} reports the two
scorings.

\begin{table}[H]
\centering
\small
\caption{\textbf{Quadratic terminal scoring captures interactions missed by
linear scoring.}  Entries are medians over five partitions and three update
budgets.  Errors are normalized as $2\varepsilon/\Delta_J$.}
\label{tab:quadratic-closure}
\begin{tabular}{lrrr}
\toprule
Domain & Linear terminal & Quadratic terminal & Residual ratio \\
\midrule
Vision & $0.145$ & $0.006$ & $0.041$ \\
AG News & $0.141$ & $0.006$ & $0.043$ \\
20 Newsgroups & $0.212$ & $0.009$ & $0.045$ \\
\bottomrule
\end{tabular}%

\end{table}

\subsection{Suffix localization from contraction}
\label{app:suffix-mechanism}

\paragraph{Exact contraction under damped Newton.}
Both frozen-ridge update rules are affine in the weights, so the Lipschitz
constants that Proposition~\ref{prop:common-suffix-memory}(a) multiplies are
not estimated.  Writing $H_t=X_r^\top X_r/n_r+\lambda I$ for the retained
Hessian and $g=H_tw-b_t$ for the gradient, $T$ gradient steps give
$M_t=(I-\eta H_t)^T$, while the damped Newton step $w\mapsto w-\delta H_t^{-1}g$
gives
\[
 M_t=I-\delta H_t^{-1}H_t=(1-\delta)I .
\]
The Hessian cancels, so at $\delta=0.5$ the ordered-suffix products are
$\kappa_1=0.5$ and $\kappa_2=0.25$ for every partition, request and suffix.
Each step's Jacobian depends only on which requests have been deleted when it
runs, and for a fixed ordered suffix that set does not depend on the prefix,
which is why the suffix map is well defined.  Measuring the realized
attenuation $\|U_s(u)-U_s(v)\|/\|u-v\|$ of every pair of prefixes entering the
same suffix reproduces these constants exactly under Newton and never exceeds
them under gradient descent, where $\kappa_1$ is $0.9947$ at one step and
$0.9190$ at sixteen.

\paragraph{Matched gradient--Newton intervention.}
Development used three instances.  The prediction was then frozen before any
held-out Newton table existed.  Two things were frozen, and they carry
different weight.  Reproducing the analytic constants above is an
implementation check.  The Jacobian is $(1-\delta)I$ by construction, so a run
disagreeing with it indicates a fault in the harness.  The prespecified
structural prediction is separate and falsifiable, and concerns the relative
movement of the suffix errors against their equal-capacity prefix and middle
controls as the contraction changes.  The bound of
Proposition~\ref{prop:common-suffix-memory}(a) also involves $R_s$ and $d_s$,
and it predicts nothing about the direction the controls take.  The five held-out partitions are
matched to the archived gradient instances on checkpoint, feature extractor,
request sample ids, $\lambda$, learning rate and step scale.  Only the
preconditioner differs.  The held-out intervention supports the prespecified
separation: stronger analytic contraction favors the suffix family relative to
its matched positional controls.  The archived gradient surfaces do not show
that separation, so the result concerns where positional memory resides rather
than a uniform reduction in family error
(Table~\ref{tab:newton-heldout}).

\begin{table}[H]
\centering
\small
\caption{\textbf{Matched gradient descent and damped Newton differ in which
positional families suffice.}  Entries are medians with observed
$[\min,\max]$ across the five held-out partitions.  Scale is
$100\Delta_J/\bar J$.}
\label{tab:newton-heldout}
\begin{tabular}{@{}lrrrrr@{}}
\toprule
Update rule & Scale (\%) & $D_{\mathcal A}$ & $D_{S,1}$ & $D_{S,2}$ & $D_{P,2}$ \\
\midrule
Gradient descent & $0.565$ & $0.003$ & $1.095$ & $0.543$ & $0.559$ \\
 & $[0.416,0.806]$ & $[0.001,0.005]$ & $[1.000,1.169]$ & $[0.496,0.579]$ & $[0.512,0.599]$ \\
Damped Newton & $31.684$ & $0.111$ & $0.597$ & $0.143$ & $1.488$ \\
 & $[29.902,47.809]$ & $[0.052,0.283]$ & $[0.555,0.768]$ & $[0.133,0.203]$ & $[1.268,1.605]$ \\
\bottomrule
\end{tabular}%

\end{table}

\paragraph{Damping sweep.}
\label{app:newton-sweep}
The damped Newton Jacobian is $(1-\delta)I$, so the one-step contraction
$\kappa_1=1-\delta$ is set by the damping and known before execution.  The
same five partitions were re-executed at three further damping values,
changing nothing else.  Those three were fixed before their outcomes were
observed, and $\delta=0.5$ had been evaluated previously.
Table~\ref{tab:newton-damping-sweep} reports every partition at every
damping.

\begin{table}[H]
\centering
\scriptsize
\setlength{\tabcolsep}{1.3pt}
\caption{\textbf{Analytically stronger Newton contraction progressively
favors suffix-localized structure.}  Entries are medians with observed
$[\min,\max]$ across the same five instances, with both scale columns in
percent and $j_0=j(w_0)$ the objective at the undeleted state.}
\label{tab:newton-damping-sweep}
\begin{tabular}{@{}rrrrrrrrr@{}}
\toprule
$\kappa_1$ & $\delta$ & $\Delta_J/\bar J$ & $\Delta_J/j_0$ & $D_{\mathcal A}$ & $D_{S,1}$ & $D_{S,2}$ & $D_{P,2}$ & $D_{M,2}$ \\
\midrule
$0.90$ & $0.10$ & \shortstack[r]{$8.263$\\$[6.563,11.721]$} & \shortstack[r]{$5.038$\\$[3.999,7.118]$} & \shortstack[r]{$0.022$\\$[0.011,0.059]$} & \shortstack[r]{$1.063$\\$[0.951,1.105]$} & \shortstack[r]{$0.487$\\$[0.431,0.495]$} & \shortstack[r]{$0.720$\\$[0.614,0.740]$} & \shortstack[r]{$1.755$\\$[1.559,1.840]$} \\
$0.75$ & $0.25$ & \shortstack[r]{$19.492$\\$[16.593,28.456]$} & \shortstack[r]{$5.769$\\$[4.890,8.202]$} & \shortstack[r]{$0.060$\\$[0.028,0.159]$} & \shortstack[r]{$0.910$\\$[0.819,1.015]$} & \shortstack[r]{$0.347$\\$[0.312,0.368]$} & \shortstack[r]{$0.991$\\$[0.810,1.016]$} & \shortstack[r]{$1.808$\\$[1.602,1.957]$} \\
$0.50$ & $0.50$ & \shortstack[r]{$31.684$\\$[29.902,47.809]$} & \shortstack[r]{$2.837$\\$[2.561,3.885]$} & \shortstack[r]{$0.111$\\$[0.052,0.283]$} & \shortstack[r]{$0.597$\\$[0.555,0.768]$} & \shortstack[r]{$0.143$\\$[0.133,0.203]$} & \shortstack[r]{$1.488$\\$[1.268,1.605]$} & \shortstack[r]{$2.051$\\$[1.742,2.182]$} \\
$0.25$ & $0.75$ & \shortstack[r]{$37.595$\\$[34.761,56.612]$} & \shortstack[r]{$0.710$\\$[0.555,0.905]$} & \shortstack[r]{$0.093$\\$[0.047,0.221]$} & \shortstack[r]{$0.247$\\$[0.222,0.408]$} & \shortstack[r]{$0.035$\\$[0.032,0.067]$} & \shortstack[r]{$1.869$\\$[1.795,2.124]$} & \shortstack[r]{$2.059$\\$[1.946,2.325]$} \\
\bottomrule
\end{tabular}%

\end{table}

The two scale columns move differently: stronger damping sharpens the
structure of the ordering problem while eventually shrinking what ordering is
worth relative to the fixed starting distance.  The intervention therefore
separates structural simplicity from decision scale rather than supporting a
claim about deployment value.

The realized contractions match their analytic values on every
instance--damping pair.  Stronger damping separates the suffix families from
the equal-capacity positional controls, whereas assignment error is
non-monotone.  Absolute suffix width is non-monotone as well, so the separation
is not an effect of the shared normalization.

\paragraph{Matched-history attenuation.}
We compare histories that delete the same two requests in opposite orders and
then apply the same two-request suffix with request-bound randomness
(Table~\ref{tab:matched-history-decay}).

\begin{table}[H]
\centering
\caption{\textbf{Matched histories contract in output space and in the
terminal objective under a common suffix.}  Statistics aggregate 120 matched
paths from ten request sets.  Output separation is RMS logit distance, and
alignment is the absolute objective difference divided by its Cauchy bound
$2\|m-z_R\|_{\rm RMS}\|d\|_{\rm RMS}$, whereas $j_{\rm RMS}$ is a mean squared
distance.}
\label{tab:matched-history-decay}
\begin{tabular}{lrrrr}
\toprule
Common suffix & Output RMS & $|j_{\rm RMS}(z)-j_{\rm RMS}(z')|$ & Cauchy bound & Alignment \\
\midrule
Before & 2.025 & 0.621 & 12.664 & 0.052 \\
After one request & 0.211 & 0.079 & 1.292 & 0.069 \\
After two requests & 0.130 & 0.045 & 0.760 & 0.058 \\
\bottomrule
\end{tabular}%

\end{table}

Let $d_z=20|\mathcal A|$ be the number of centered-logit coordinates and define
$j_{\rm RMS}(z)=d_z^{-1}\|z-z_R\|_2^2$.  The main full-network objective is
$j(z)=20j_{\rm RMS}(z)$: it averages squared Euclidean distance across audit
examples but sums over the 20 superclass coordinates.  This constant factor
does not affect attenuation ratios or alignment, while the absolute objective
differences and bounds in Table~\ref{tab:matched-history-decay} use
per-coordinate MSE units.
For $d=z-z'$ and $m=(z+z')/2$, the exact identity
$j_{\rm RMS}(z)-j_{\rm RMS}(z')=(2/d_z)\langle m-z_R,d\rangle$ holds on every pair.  From the initial
prefix to the completed common suffix, median output retention is $6.62\%$ and
median absolute-objective-difference retention is $6.69\%$.  Their agreement,
together with the stable alignment fraction, attributes the smaller objective
width primarily to output attenuation.  Because $d$ is observed, this is a
finite-execution consistency analysis.  Run on the three matched pilots per
policy that store prefix states, the same comparison leaves a median
objective-difference retention of $6.9\%$ under NegGrad+, observed range
$[6.5,7.2]$, and $10.2\%$ under SalUn, observed range $[10.1,10.5]$.  The next
experiment evaluates held-out
transitions.

\paragraph{Conditioning versus truncation.}
Proposition~\ref{prop:common-suffix-memory} states two rates at a fixed suffix
depth, $O(\rho^L)$ for truncation and $O(\rho^{2L})$ for conditioning, and this
is where the two are measured against each other on executed surfaces.  The
comparison needs a design the conditional family cannot saturate.  At $k=4$ with
$L=2$ each suffix leaves two prefix orders that an assignment fits exactly, so
an error of zero there would report capacity.  We therefore use the complete
$k=5$ NegGrad+ surfaces at $L=1$, where each of the five suffixes leaves $24$
prefix orders against ten free parameters, $50$ in rank across the $120$ orders.

Conditioning is more accurate than either nested comparator on every instance.
This checks the construction at one depth on a design where the conditional
family has ample room to miss.  Establishing the exponent would require varying
$L$ or $\rho$.
Because the three nested complete-table families have unequal rank, the main
text separately compares first-, third-, and final-position conditioning at
matched rank 50.  Its four folds are fixed from cyclic relabeling orbits before
outcomes are read; every 90-order training design has full rank and balanced
conditioned-request counts.  For fold $f$, define the training-only center and
scale
\[
 \overline J_f^{\rm train}
 =\frac{1}{|T_f^{\rm train}|}\sum_{\pi\in T_f^{\rm train}}J(\pi),
 \qquad
 \Delta_f^{\rm train}
 =\overline J_f^{\rm train}-\min_{\pi\in T_f^{\rm train}}J(\pi),
\]
and $\widetilde J_f^{\rm train}(\pi)=
(J(\pi)-\overline J_f^{\rm train})/\Delta_f^{\rm train}$.  Each family
$\mathcal F_r$ is fitted by
\[
 \widehat g_{r,f}\in\arg\min_{g\in\mathcal F_r}
 \max_{\pi\in T_f^{\rm train}}
 |\widetilde J_f^{\rm train}(\pi)-g(\pi)|,
 \qquad
 \widehat J_{r,f}(\pi)=\overline J_f^{\rm train}
 +\Delta_f^{\rm train}\widehat g_{r,f}(\pi).
\]
If the minimax solution is nonunique, a pre-fixed tie rule minimizes the
prediction norm $\sum_{\pi\in S_5}g(\pi)^2$; this uses the complete permutation
design but no held-out outcome.  Only evaluation uses the complete-surface
headroom $\Delta_J$:
\[
 D_{r,f}^{\rm test}=\frac{2}{\Delta_J}\max_{\pi\in T_f}
 |J(\pi)-\widehat J_{r,f}(\pi)|.
\]
Thus the fitted original-scale predictions depend only on the training
outcomes.  The final-position model wins both controls in every instance--fold
comparison.  This supports positional specificity, not the affine decay
exponent.

\subsection{Separation of structural families}
\label{app:family-separation}

\paragraph{Nested position families.}
Table~\ref{tab:sorting-versus-assignment} compares three nested position
families.  The linear family fixes the shared positional profile, the monotone
family learns that profile jointly with one score per request, and assignment
allows an unrestricted request--position cost matrix.  The monotone fit is
retrospective and non-convex.  The reported value is the best of the two
search procedures fixed in advance, which agree to $5\times10^{-5}$ on
every
surface.

\begin{table}[H]
\centering
\small
\caption{\textbf{Nested position families separate fixed-profile sorting,
learned-profile sorting, and unrestricted assignment.}  $D_{\rm lin}$ fixes a
shared linear positional profile, $D_{\rm mono}$ learns a shared monotone
profile, and $D_{\mathcal A}$ is unrestricted.  $R$ is the normalized regret of
the selected order.  The last column counts instances meeting, on, and failing
the criterion for $D_{\rm mono}$, with the boundary band $[0.95,1.05]$.}
\label{tab:sorting-versus-assignment}
\begin{tabular}{lrrrrrrr}
\toprule
Setting & Instances & $D_{\rm lin}$ & $D_{\rm mono}$ & $D_{\mathcal A}$ & $R_{\rm mono}$ & $R_{\mathcal A}$ & m/b/f \\
\midrule
Frozen ridge, vision & 15 & $0.056$ & $0.011$ & $0.010$ & $0.000$ & $0.000$ & 15/0/0 \\
Frozen ridge, AG News & 15 & $0.055$ & $0.026$ & $0.018$ & $0.000$ & $0.000$ & 15/0/0 \\
Frozen ridge, 20 Newsgroups & 15 & $0.061$ & $0.055$ & $0.022$ & $0.000$ & $0.000$ & 15/0/0 \\
Direct softmax head & 15 & $0.071$ & $0.068$ & $0.035$ & $0.000$ & $0.000$ & 14/0/1 \\
Factorized softmax head & 15 & $0.545$ & $0.305$ & $0.065$ & $0.009$ & $0.000$ & 13/0/2 \\
NegGrad+ & 10 & $1.582$ & $0.281$ & $0.236$ & $0.122$ & $0.064$ & 10/0/0 \\
SalUn & 10 & $1.346$ & $0.857$ & $0.673$ & $0.348$ & $0.256$ & 6/1/3 \\
\bottomrule
\end{tabular}%

\end{table}

The learned profile nearly closes the gap to assignment for NegGrad+, whereas
SalUn exhibits substantially greater instance-to-instance variation.  The
apparent failure of scalar sorting under a fixed linear profile therefore does
not extend to the more general shared-profile family.

\paragraph{Matched parameterization intervention.}
\label{app:primary-structure-tables}
We compare direct logistic regression with its function-matched factorization
$W=VU$ on held-out request unions.  All configurations passed the frozen
validity criteria.  Factorization
increased the paired order range in 14/15 instances and changed the optimal order
in 8/15.  Assignment met $D_{\mathcal A}<1$ in every instance and was substantially
tighter than scalar sorting.  The initial predictor and executed objective
were matched, so the contrast isolates the update parameterization.

\paragraph{Within-suffix interaction.}
\label{app:suffix-additivity}
The suffix criterion can be met for two different reasons.  Either the final
two positions contribute additively, in which case the suffix structure is the
tail of an assignment cost matrix, or the two surviving requests interact.
Table~\ref{tab:suffix-additivity} separates them by restricting the same
family to $c+u_{\pi_{k-1}}+v_{\pi_k}$.  The additive family is contained in
the general one, so the difference is exactly the within-suffix interaction,
and it is contained in assignment, so $D_{\mathcal A}$ lower-bounds it.

\begin{table}[H]
\centering
\small
\caption{\textbf{General suffix localization can retain interactions that an
additive suffix cannot represent.}  Entries are medians across instances with observed
$[\min,\max]$ on complete full-network permutation surfaces.  All suffix fits
condition on the final two requests.}
\label{tab:suffix-additivity}
\begin{tabular}{@{}lrrrr@{}}
\toprule
Setting & Instances & $D_S$ & $D_S^{\rm add}$ & $D_{\mathcal A}$ \\
\midrule
NegGrad+ & 10 & \shortstack[r]{$0.148$\\$[0.055,0.477]$} & \shortstack[r]{$0.293$\\$[0.146,0.489]$} & \shortstack[r]{$0.236$\\$[0.119,0.391]$} \\
NegGrad+ ($k=5$) & 3 & \shortstack[r]{$0.238$\\$[0.227,0.582]$} & \shortstack[r]{$0.238$\\$[0.230,0.924]$} & \shortstack[r]{$0.186$\\$[0.150,0.771]$} \\
SalUn & 10 & \shortstack[r]{$0.260$\\$[0.185,0.510]$} & \shortstack[r]{$0.746$\\$[0.335,1.139]$} & \shortstack[r]{$0.673$\\$[0.215,1.030]$} \\
\bottomrule
\end{tabular}%

\end{table}

SalUn shows a material gap between general and additive suffix fits, unlike
NegGrad+.  Its localized dependence therefore retains interactions that cannot
be represented as separate costs for the final two positions.

\paragraph{Equal-capacity positional controls.}
To test whether suffix fits are favored merely by flexibility, we compare them
with arbitrary functions of the requests occupying the first $L$ positions or
a fixed contiguous middle block.  Each family has
$k!/(k-L)!$ groups, and all widths use the same $\Delta_J$ normalization
(Table~\ref{tab:position-memory-control-details}).

\begin{table}[H]
\centering
\caption{\textbf{Equal-capacity positional controls distinguish suffix
localization from generic low-dimensional dependence.}  Prefix, middle, and
suffix families condition on the same number of ordered positions and use the
same $\Delta_J$ normalization.}
\label{tab:position-memory-control-details}
\begin{tabular}{lrrrr}
\toprule
Instances & $L$ & Prefix & Middle & Suffix \\
\midrule
$k=4$ (10) & 1 & 2.137 & 2.192 & 0.368 \\
 & 2 & 1.926 & 2.096 & 0.148 \\
$k=5$ (3) & 1 & 2.254 & 2.244 & 0.331 \\
 & 2 & 2.232 & 2.243 & 0.238 \\
\bottomrule
\end{tabular}%

\end{table}

The matched controls favor suffix localization overall, although less
uniformly than the damped-Newton intervention.  This comparison isolates
recency from generic conditioning capacity but does not construct an online
suffix score.

\paragraph{Queue-length extension at $k=5$.}
The extension keeps the NegGrad+ policy fixed and changes only the number of
superclass requests from four to five.  Each seed contains all 120 orders and
325 nonempty prefix executions, and
Table~\ref{tab:k5-scale} gives the resulting widths.

\begin{table}[H]
\centering
\caption{\textbf{At $k=5$, the general suffix error separates prefix memory
from additive suffix structure.}  All entries are normalized by the total ordering
gap $\Delta_J$.}
\label{tab:k5-scale}
\begin{tabular}{lrrrr}
\toprule
Seed & Suffix $L{=}1$ & General suffix $L{=}2$ & Additive suffix $L{=}2$ & Assignment \\
\midrule
0 & 0.327 & 0.227 & 0.230 & 0.186 \\
1 & 1.005 & 0.582 & 0.924 & 0.771 \\
2 & 0.331 & 0.238 & 0.238 & 0.150 \\
\bottomrule
\end{tabular}%

\end{table}

The extension supports suffix localization while leaving room for
within-suffix interactions.  The strict separation is supplied by
Proposition~\ref{prop:memory-additivity-separation}; this experiment does not
estimate a contraction parameter.

\subsection{Nonlinear reach of the suffix bound}
\label{app:nonlinear-reach}

\paragraph{Direct-head analytic bound.}
Proposition~\ref{prop:common-suffix-memory}(a) is the one quantitative result
that covers updates which are not affine, so how much it delivers on an
executed nonlinear head sets what we are entitled to say about the
full-network policies.  The direct head appends a constant feature for bias and
regularizes the entire two-class weight matrix, including that coordinate.
Each deletion transition uses one, four, or sixteen ordinary full-batch SGD
steps with learning rate $0.05$, mean cross-entropy, $\lambda=10^{-4}$, and
fresh optimizer state.  For a retained design $X$, the softmax Hessian
therefore lies between $\lambda I$ and $(\lambda+\|X\|_2^2/(2n))I$.  We compute
the resulting contraction factor separately for the two retained sets in every
ordered suffix and use the exact operator norm of the validation design as
$L_\Phi$.

The comparison uses every one of the 15 matched direct-head instances.  None is
selected by its gap or by bound tightness.  Median analytic two-stage
contraction factors are $0.99999$, $0.99996$, and $0.99984$ across the three
budgets, and the corresponding finite-pair endpoint ratios are $0.996$, $0.978$,
and $0.919$.  The latter are trajectory diagnostics rather than regional
Lipschitz constants.  The global objective bound holds for every suffix in
every instance, and Table~\ref{tab:direct-nonlinear-bound} reports how
conservative it is on absolute and decision-normalized scales.

The analytic bound remains valid but conservative on the executed nonlinear
heads.  We therefore use it as a sufficient-condition check, ground the
mechanism claim in the matched Newton intervention, and treat full-network
attenuation as finite-execution evidence.

\begin{table}[H]
\centering
\small
\caption{\textbf{The analytic common-suffix bound holds on executed direct-head
trajectories with roughly an order of magnitude to spare.}  Entries are medians
over five partitions.  Curvature constants are global to each retained design,
while reachable-state diameters and target radii are evaluated on the executed
instances.  $\mathcal B_2$ is the bound of
Equation~\ref{eq:nonlinear-common-suffix-bound}.}
\label{tab:direct-nonlinear-bound}
\begin{tabular}{rrrrrr}
\toprule
Steps & $W_2$ & $\mathcal B_2$ & $\mathcal B_2/W_2$ & $D_S$ & $\mathcal B_2/\Delta_J$ \\
\midrule
1 & $2.39{\times}10^{-8}$ & $1.68{\times}10^{-7}$ & 9.24 & 0.623 & 5.85 \\
4 & $8.84{\times}10^{-8}$ & $6.66{\times}10^{-7}$ & 9.78 & 0.614 & 6.05 \\
16 & $2.65{\times}10^{-7}$ & $2.51{\times}10^{-6}$ & 11.93 & 0.575 & 6.86 \\
\bottomrule
\end{tabular}%

\end{table}

\section{Out-of-sample structural identification}
\label{app:identification}

This appendix asks whether structural families identified from limited
information continue to predict orders that were not used to construct them.
The budgets run from no terminal-order queries at all, through bases of size
$O(k)$ and $O(k^2)$, to a balanced 112-query full-network design.
Table~\ref{tab:structural-regimes-sampled} carries the structural diagnostics
for the two settings whose order space is too large to enumerate.

\subsection{Probe-based structural reconstruction}
\label{app:probe-reconstruction}

The lowest budget uses no terminal-order queries at all.  The constructions in
Figure~\ref{fig:structural-axes}b read the $k$ singleton responses and the
retained derivatives, and their values are reported on queues held out of the
earlier retained-loss development.  The unary quadratic construction carries no
cross-request interaction and is the one the interaction term separates from
(Table~\ref{tab:probe-reconstruction-calibration}).  The comparative ordering
of the four remains the same on the held-out queues, although their numerical
values also change.  Fitting the
assignment family directly to the complete terminal-objective table instead of
to singleton responses gives $0.019$ and $0.017$, so the response-level
construction is the more accurate of the two here despite seeing no order.
Errors are instance medians, only an order-independent constant is removed
before aligning them, and no score scale is refitted.  The comparison is
reference-assisted and retrospective.

\begin{table}[H]
\centering
\small
\caption{\textbf{The four singleton-response constructions on the calibration
queues.}  Medians across instances of $2\varepsilon/\Delta_J$, with the mean
normalized regret of the order each construction selects.
Figure~\ref{fig:structural-axes}b reports the same constructions on the
fresh-decision queues.}
\label{tab:probe-reconstruction-calibration}
\begin{tabular}{@{}lrr@{}}
\toprule
Construction & Reconstruction error & Selection regret \\
\midrule
Linear & $0.151$ & $0.0051$ \\
Unary quad. & $0.098$ & $0.0005$ \\
Projected quad. & $0.021$ & $0.0001$ \\
Full quad. & $0.006$ & $0.0000$ \\
\bottomrule
\end{tabular}%

\end{table}

\begin{table}[H]
\centering
\small
\caption{\textbf{The same diagnostics on the two sampled designs.}  Entries are
medians across instances with observed $[\min,\max]$.  Every error is fitted on
the identification orders and evaluated on the declared held-out split, so it
describes that split rather than $S_k$.  The $k=15$ design carries no suffix
axis because $W_2$ was not estimated there.}
\label{tab:structural-regimes-sampled}
\begin{tabular}{@{}llrrrrr@{}}
\toprule
Setting & Design & Req. & Inst. & Gap (\%) & $D_{\mathcal A}^{\rm test}$ & $D_S^{\rm test}$ \\
\midrule
Ridge head & 197/500 & 15 & 3 & \shortstack[r]{$2.651$\\$[1.835,2.785]$} & \shortstack[r]{$0.110$\\$[0.101,0.160]$} & --- \\
NegGrad+ & 112/336 & 8 & 3 & \shortstack[r]{$11.886$\\$[11.024,16.399]$} & \shortstack[r]{$0.446$\\$[0.392,0.520]$} & \shortstack[r]{$0.570$\\$[0.551,0.713]$} \\
\bottomrule
\end{tabular}%

\end{table}

\subsection{Large-queue position-structure identification}
\label{app:quadratic-ablation}

\paragraph{Large-$k$ structural identification.}
The large-queue study uses three fixed CIFAR-10 request partitions and the
one-step retained-gradient policy with an exact retained reference.  Pivoted QR
selects the assignment basis before any objective value is observed.  Because
the basis is saturated, all structural evaluation uses disjoint orders, the
remaining complete surface at $k=8$ and a sampled set at $k=15$, against a
capacity-matched shuffle on the same evaluated orders and objective values
(Table~\ref{tab:ridge-identification-audit}).

Assignment generalizes beyond its saturated basis, whereas the matched shuffle
does not.  Table~\ref{tab:k8-global-rank-audit} separately audits the order
returned by solving each fitted $k=8$ model against the complete surface.  The
recovery comes from extrapolating the fitted family rather than observing the
returned order in the identification basis.

\begin{table}[H]
\centering
\small
\setlength{\tabcolsep}{4pt}
\caption{\textbf{Identification budgets and held-out fit for the two queue
lengths.}  Medians across the three partitions.  The held-out error is
$2\varepsilon/\Delta$ on the evaluation orders, and the last three columns
describe the assignment fit, which the sorting tiers do not report.  Shuffle
$R^2$ is the capacity-matched permutation of the same objective values.  The
$k$-order and assignment rows are the ones
Figure~\ref{fig:fullnet-identification}a carries.}
\label{tab:ridge-identification-audit}
\begin{tabular}{@{}rlrrrrrr@{}}
\toprule
Req. & Basis & Orders & Held-out error & Evaluated & $R^2$ & NRMSE & Shuffle $R^2$ \\
\midrule
8 & Sorting, $k$ & 8 & $0.068$ & --- & --- & --- & --- \\
8 & Sorting, $k{+}1$ & 9 & $0.066$ & --- & --- & --- & --- \\
8 & Assignment & 50 & $0.055$ & 40{,}270 & $0.9997$ & $0.017$ & $-3.69$ \\
\addlinespace[2pt]
15 & Sorting, $k$ & 15 & $0.107$ & --- & --- & --- & --- \\
15 & Sorting, $k{+}1$ & 16 & $0.094$ & --- & --- & --- & --- \\
15 & Assignment & 197 & $0.110$ & 500 & $0.9971$ & $0.054$ & $-10.61$ \\
\bottomrule
\end{tabular}%

\end{table}

\paragraph{Sorting bases for the same surfaces.}
The sorting comparison in Figure~\ref{fig:fullnet-identification}a chooses its
identification orders by the same pivoted-QR rule applied to the sorting design
alone.  The basis is drawn from $S_8$ at $k=8$, and at $k=15$, where no
complete surface exists,
from the $197$ executed identification orders, which asks whether fifteen of
those queries would have served.  At $k=8$, the fitted score is minimized over
all of $S_8$ and audited on the complete surface.  At $k=15$, it is minimized
only over the disjoint 500-order evaluation sample, so the reported rank is not
a global statement over $S_{15}$.  The extra basis order does not change the
qualitative comparison, while assignment remains the tighter description on
the complete $k=8$ surface.

\begin{table}[H]
\centering
\small
\setlength{\tabcolsep}{3.5pt}
\caption{\textbf{True rank of the orders returned by the fitted $k=8$
families.}  Each fitted score is minimized over the complete permutation space;
regret and uniqueness are then evaluated on the stored surface.}
\label{tab:k8-global-rank-audit}
\begin{tabular}{rlrrrc}
\toprule
Instance & Family & Queries & True rank & $R_{\rm cand}$ & Unique \\
\midrule
0 & Sorting & 8 & 1/40,320 & 0 & Yes \\
0 & Assignment & 50 & 1/40,320 & 0 & Yes \\
1 & Sorting & 8 & 1/40,320 & 0 & Yes \\
1 & Assignment & 50 & 1/40,320 & 0 & Yes \\
2 & Sorting & 8 & 1/40,320 & 0 & Yes \\
2 & Assignment & 50 & 1/40,320 & 0 & Yes \\
\bottomrule
\end{tabular}%

\end{table}

After fitting, we also solved each $k=15$ assignment globally with the Hungarian
algorithm and executed that one predicted order.  None of the three orders had
been evaluated during identification or sampled testing
(Table~\ref{tab:k15-global-assignment}).

\begin{table}[H]
\centering
\small
\caption{\textbf{The fitted $k=15$ assignment minimizer improves on both
identification and sampled evaluation orders when executed.}  Values are
squared distances to the exact retained ridge optimum.  Improvement is measured
relative to the mean sampled held-out objective.}
\label{tab:k15-global-assignment}
\begin{tabular}{rrrrr}
\toprule
Partition & Selected & Best identification & Best sampled & Improvement \\
\midrule
0 & $2.2670\!\times\!10^{-4}$ & $2.2880\!\times\!10^{-4}$ & $2.2826\!\times\!10^{-4}$ & $3.32\%$ \\
1 & $2.2730\!\times\!10^{-4}$ & $2.2873\!\times\!10^{-4}$ & $2.2796\!\times\!10^{-4}$ & $3.07\%$ \\
2 & $2.2877\!\times\!10^{-4}$ & $2.3055\!\times\!10^{-4}$ & $2.3023\!\times\!10^{-4}$ & $2.46\%$ \\
\bottomrule
\end{tabular}%

\end{table}

\subsection{Full-network held-out prediction}
\label{app:fullnet-heldout}

\paragraph{Balanced sampled design at $k=8$.}
We used a balanced 112/336 identification/held-out design in which every
ordered request pair occurs equally often at each tested position pair.  The
nonoverlapping order sets are disjoint orbits of a fixed affine permutation
group.  This construction gives exact request-pair balance at each tested
position pair.

For each position-pair family, the value of a request-pair group was the
midpoint of its two identification executions.  The assignment coefficients were
instead obtained by a minimax fit using only the same 112 identification
executions.  No coefficient, offset, family, or orientation was adjusted using
the held-out executions.  All seeds used the same order split and request
partition.  Three seeds changed the model/execution instance.  For a fitted
family $\mathcal F$ with surrogate $\widehat J_{\mathcal F}$, the held-out error
is
\[
 D_{\mathcal F}^{\rm test}
 =\frac{2\max_{\pi\in\mathcal T}
 |J(\pi)-\widehat J_{\mathcal F}(\pi)|}
 {\Delta_{\rm sample}},
\]
where $\mathcal T$ is the held-out set and $\Delta_{\rm sample}$ is the
mean-to-minimum gap across all 448 sampled orders.  This denominator is an
evaluation scale, not an input to fitting.  Each of the three 448-order seed
tables passed the validity screen on all sampled orders.

All three positional families condition on two ordered positions and therefore
have equal capacity.  Their held-out comparison is reported in
Table~\ref{tab:k8-sampled-suffix}.

\begin{table}[H]
\centering
\small
\caption{\textbf{At $k=8$, assignment and suffix structure generalize to
held-out full-network orders while equal-capacity controls do not.}  Each
instance uses 112 identification and 336 held-out orders with exact pair
balance.  Reported errors are evaluated only on the held-out design, and the
last column is the observed suffix error, the width normalized by the
sampled gap.}
\label{tab:k8-sampled-suffix}
\begin{tabular}{lrrrrr}
\toprule
Instance & Prefix & Middle & Suffix & Assignment & Observed $D_S$ \\
\midrule
0 & 2.775 & 3.578 & 0.551 & 0.392 & 0.300 \\
1 & 2.592 & 2.743 & 0.570 & 0.520 & 0.376 \\
2 & 4.505 & 4.505 & 0.713 & 0.446 & 0.418 \\
\bottomrule
\end{tabular}%

\end{table}

\subsection{Validation on unseen executions}
\label{app:fullnet-candidates}

\paragraph{Candidate transfer.}
We fixed each fitted family and its tie-breaking rule before executing the
returned candidates: the original 112 identification executions, the minimax
assignment and suffix fits, lexicographic tie breaking, and an ascending
request-ID completion for the unspecified prefix.  Executing the two
candidates per instance under the original request-bound randomness, all six
passed the validity screen (Table~\ref{tab:k8-candidate-execution}).

Candidate quality is mixed across both families and instances, so these runs do
not establish a stable search advantage at matched budget.  Their role is to
verify that a model fitted on one set of executions can produce a valid unseen
candidate; the structural comparison remains the held-out prediction result in
Figure~\ref{fig:fullnet-identification}.

\begin{table}[H]
\centering
\small
\caption{\textbf{Executed candidates against the fixed $448$-order design.}
Rank $1$ is the best order in that pool.  Regret is normalized by the sampled
gap, so a negative value means the candidate beats every order in the pool.
The last column is the exact probability that 113 draws without replacement
from this finite design match or beat the candidate; it is not a probability
under uniform search over $S_8$.}
\label{tab:k8-candidate-execution}
\begin{tabular}{@{}rlrrr@{}}
\toprule
Instance & Family & Rank & Regret & Random-search match \\
\midrule
0 & Assignment & 20 & $+0.121$ & $0.997$ \\
0 & Suffix & 19 & $+0.114$ & $0.995$ \\
1 & Assignment & 2 & $+0.013$ & $0.252$ \\
1 & Suffix & 13 & $+0.116$ & $0.971$ \\
2 & Assignment & 1 & $-0.038$ & $0.000$ \\
2 & Suffix & 46 & $+0.129$ & $1.000$ \\
\bottomrule
\end{tabular}%

\end{table}

\section{Additional controls and diagnostics}
\label{app:controls}

This appendix asks whether the structure reported in
Appendices~\ref{app:structural-evidence} and~\ref{app:identification}
could instead be explained by family capacity, by the choice of terminal
objective, or by the normalization the errors use.

\subsection{Capacity controls}
\label{app:capacity-controls}

\paragraph{Capacity-matched nulls.}
For each head instance, we shuffled the 24 terminal objectives
across permutations 1,000 times and refit the same minimax assignment family.
The null leaves the objective multiset, gap, and assignment design unchanged.
We evaluate both separation from this capacity-matched null and the realized
objective regret of the fitted assignment optimizer
(Table~\ref{tab:assignment-capacity-audit}).  The latter is not
bounded tightly by $D_{\mathcal A}<1$ alone.

\begin{table}[H]
\centering
\caption{\textbf{Assignment structure exceeds a capacity-matched
shuffled-label baseline and usually selects a near-optimal order.}
``Optimal'' counts instances whose fitted assignment minimizer attains the
terminal-objective minimum.  Regret is
$(J(\widehat\pi)-J(\pi^\star))/\Delta_J$.}
\label{tab:assignment-capacity-audit}
\begin{tabular}{lrrrr}
\toprule
Setting & Instances & Shuffle $p$ & Optimal & Max. regret \\
\midrule
Frozen ridge & 45 & $\le 0.001$ & 44/45 & $0.0007$ \\
Direct head & 15 & $\le 0.001$ & 13/15 & $0.151$ \\
Factorized head & 15 & $\le 0.001$ & 10/15 & $0.452$ \\
\bottomrule
\end{tabular}%

\end{table}

We permuted each table's objective labels 1,000 times and re-fit the identical
suffix $L{=}1$, suffix $L{=}2$, and assignment families, which retains the objective
distribution and gap while destroying the request--position association.
On the exhaustive surfaces all three families beat every one of the 1,000
shuffles in all ten instances and all three at $k=5$.  None of the $1{,}000$
capacity-matched shuffles matched the observed errors.

The separation from the shuffled fits shows that the observed localization
contains request--position information beyond family capacity.

\subsection{Behavioural anchoring of the terminal objective}
\label{app:behavioral-anchoring}

The terminal objective is a dense structural-analysis objective, chosen so that
the induced permutation problem is measurable on every order.  This subsection
asks what accompanies the ordering variation it exposes, using the executions
already reported and adding none.  The complete $k=4$ full-network surfaces
record retained and deleted-group accuracy for each of the $24$ orders, so both
quantities are available beside each objective value.

Every quantity below is measured within one policy and one instance.  The two
policies optimise different things, so an order effect under one of them
describes that policy alone, and the table places their accuracies on a common
scale only to the extent of asking whether an effect is present.  The contrasts
are differences from the random-order mean of the same surface, and the rank
correlation is Spearman's between the $24$ objective values and the $24$
accuracies of one instance (Table~\ref{tab:behavioral-anchoring}).

Retained accuracy is more closely aligned with the objective than deleted-group
accuracy under both policies.  The latter varies more across orders but has
little rank alignment with the objective.  Terminal reference distance is
therefore a utility-anchored comparison objective, not a forgetting metric; a
deployment would pair it with a separate forgetting criterion.

\begin{table}[H]
\centering
\small
\setlength{\tabcolsep}{4pt}
\caption{\textbf{Ordering variation in the terminal objective is accompanied by
a consistent retained-accuracy effect within each policy.}  Entries are medians
with observed $[\min,\max]$ over ten complete $k=4$ surfaces per policy.  The
first two columns are percentage-point differences from the random-order mean
of the same surface, the third is the observed spread across its $24$ orders,
and the last is the within-instance Spearman correlation between objective and
accuracy.  Rows describe one policy at a time.}
\label{tab:behavioral-anchoring}
\begin{tabular}{llrrrr}
\toprule
Policy & Quantity & $J$-best & $J$-worst & Range & Rank corr. \\
\midrule
NegGrad+ & Retained acc. & $+0.603$ & $-0.569$ & $1.913$ & $-0.670$ \\
 & Deleted-group acc. & $+0.583$ & $+0.660$ & $4.533$ & $-0.166$ \\
\addlinespace[2pt]
SalUn & Retained acc. & $+0.772$ & $-0.711$ & $2.189$ & $-0.833$ \\
 & Deleted-group acc. & $+0.502$ & $-0.801$ & $5.397$ & $-0.087$ \\
\bottomrule
\end{tabular}%

\end{table}

Absolute levels and the screens they are compared with are reported in
Table~\ref{tab:fullnet-behavior-levels}.  The deleted-group screen requires the
largest terminal $A_F$ over all orders to be at most
$A_F^{\rm base}-0.05$ and is checked retrospectively.  The retained screen
requires the smallest terminal retained accuracy to exceed
$\min(A_R^{\rm base},A_R^{\rm retrain})-0.05$.  Reference progress requires
the order-mean terminal distance to be below the base-model distance and was
used, with the retained screen, in operating-point selection.  These checks
are not a privacy guarantee.

\begin{table}[H]
\centering
\small
\setlength{\tabcolsep}{3.5pt}
\caption{\textbf{Absolute deleted-group accuracy and declared screens.}
Entries are medians over ten complete $k=4$ surfaces per policy.  The bracket
in $A_F^{\rm unlearn}$ is the range over every terminal order and instance;
$d_U/d_0$ is the order-mean reference distance divided by the base distance.
The $F$ screen and $R$ floor columns report the median instance thresholds.}
\label{tab:fullnet-behavior-levels}
\begin{tabular}{lrrrrrr}
\toprule
Policy & $A_F^{\rm base}$ & $A_F^{\rm retrain}$ & $A_F^{\rm unlearn}$ & $F$ screen & $R$ floor & $d_U/d_0$ \\
\midrule
NegGrad+ & $0.893$ & $0.000$ & $0.695$ $[0.569,0.835]$ & $0.843$ & $0.848$ & $0.617$ \\
SalUn & $0.900$ & $0.000$ & $0.737$ $[0.613,0.839]$ & $0.850$ & $0.848$ & $0.708$ \\
\bottomrule
\end{tabular}%

\end{table}

The unlearned deleted-group accuracies remain well above their retraining
counterparts.  The screens establish a decrease, retained utility, and
reference progress at the tested operating points; they do not establish
retraining-equivalent forgetting.

\subsection{Robustness to execution randomness}
\label{app:randomness-robustness}

Repeating a fixed order under the same request-bound tape reproduces the
terminal objective exactly, so the question here is not numerical
reproducibility but whether the ordering signal survives a different draw.  We
paired two separate replays of all $24$ orders for three $k=4$ NegGrad+
request sets, holding the model, requests, reference, budget, and 10,000-example
audit fixed while changing the request-bound tape
(Table~\ref{tab:randomness-robustness}).

The two tapes select the same best order in only one of the three instances,
whereas their assignment and suffix verdicts agree.  Thus the exact ranking is
less stable than the supported structural family.  Only NegGrad+ was paired in
this way, so this diagnostic does not establish the same behavior for SalUn or
coefficient-level stability across tapes.

\begin{table}[H]
\centering
\small
\caption{\textbf{A different request-bound tape moves individual objective
values but not the structural verdict.}  Entries are medians with observed
$[\min,\max]$ over three paired complete $k=4$ NegGrad+ surfaces.  The first
block is measured between the two tapes, normalized by the random-order
gap.  The second reports each family's error under each tape separately.}
\label{tab:randomness-robustness}
\begin{tabular}{@{}lr@{}}
\toprule
Quantity & Median $[\min,\max]$ \\
\midrule
RMS objective disagreement & \shortstack[r]{$0.193$\\$[0.180,0.270]$} \\
Largest objective disagreement & \shortstack[r]{$0.409$\\$[0.357,0.519]$} \\
Rank correlation between tapes & \shortstack[r]{$0.867$\\$[0.862,0.945]$} \\
\addlinespace[2pt]
$D_{\mathcal A}$, first tape & \shortstack[r]{$0.302$\\$[0.265,0.441]$} \\
$D_{\mathcal A}$, second tape & \shortstack[r]{$0.267$\\$[0.189,0.347]$} \\
$D_{S,2}$, first tape & \shortstack[r]{$0.221$\\$[0.204,0.286]$} \\
$D_{S,2}$, second tape & \shortstack[r]{$0.204$\\$[0.126,0.276]$} \\
\bottomrule
\end{tabular}%

\end{table}

\subsection{Relation to existing ordering rules}
\label{app:existing-rules}

\citet{kumar2026controlling} order a forget set by base-model predictive
uncertainty.  That rule acts on the object this paper permutes, a fixed request
set whose order is the decision, so it can be placed on our surfaces directly.
\citet{zhao2024what} instead vary how a forget set is decomposed into
sequential requests.  That changes the request set rather than the order of
a fixed one, so it is not a permutation rule on the same $S_k$, and we do not
treat it as one here.

The uncertainty rule was executed reference-free on ten full-network
instances, reading one base-model forward pass per request, loading no
retraining reference and issuing no terminal-order query.  Those ten instances also have
complete $24$-order surfaces, so the order it selects can be scored exactly
without executing anything further
(Table~\ref{tab:kumar-rule-on-surfaces}).

The uncertainty rule neither selects competitive orders consistently nor
tracks the scalar score induced by the terminal-distance objective.  This is a
matched comparison on the same request sets, but the rule was designed for a
different objective.  Its reference-free information advantage therefore
motivates the fixed-objective audit in the next subsection rather than a claim
that uncertainty ordering fails for its intended use.

\begin{table}[H]
\centering
\small
\caption{\textbf{An existing uncertainty rule, scored exactly on the complete
surfaces.}  Ten complete $k=4$ NegGrad+ surfaces, the same instances on which
the rule was executed reference-free.  Rank $1$ is the best of the $24$ orders,
regret is normalized by the random-order gap so that $1$ is the random-order
mean, and the correlation is Spearman's between the rule's per-request scores
and the scalar sorting score of Proposition~\ref{prop:slice-additive-certificate}(b).}
\label{tab:kumar-rule-on-surfaces}
\begin{tabular}{@{}lr@{}}
\toprule
Quantity & Median $[\min,\max]$ \\
\midrule
Rank among the $24$ orders & 15 $[7, 24]$ \\
Normalized regret & \shortstack[r]{$1.455$\\$[0.177,2.999]$} \\
Better than the random-order mean & 3/10 \\
Correlation with the induced score & \shortstack[r]{$0.300$\\$[-0.400,1.000]$} \\
\bottomrule
\end{tabular}%

\end{table}

\subsection{Structure under reference-free terminal objectives}
\label{app:reference-free}

The terminal objective compares an executed model with a retraining reference,
which a deployment does not have.  We therefore used three separate 12-step
replays with the same seed-indexed requests and reference files as the first
three main-cohort instances, but evaluated on the full 10,000-example test
split.  On each replay we recomputed the structure under two fixed objectives
that need no retraining.  Let $c_y(x;w)$ be the terminal softmax probability of
the true label.  For request $i$, let $\widehat F_i^{\rm del}$ be the empirical
CDF of this score on its deleted training examples and let
$\widehat F_i^{\rm nm}$ use fixed CIFAR-100 test examples from the same
superclass.  With $q_b=b/10$, $b=1,\ldots,9$, we define
\[
 J_{\rm NM}(w)=\sum_i\sum_{b=1}^9
 \bigl(\widehat F_i^{\rm del}(q_b)-
       \widehat F_i^{\rm nm}(q_b)\bigr)^2.
\]
The second objective is the policy loss on fixed sets,
\[
 J_{\rm pol}(w)=0.95\,\mathrm{CE}(w;R_{\rm eval})
                -0.05\,\mathrm{CE}(w;F_{\rm union}),
\]
where $R_{\rm eval}$ is one fixed 5,000-example permanent-retain sample and
$F_{\rm union}$ is the union of all four deleted groups.  The same examples,
thresholds, and weights are used for every permutation.  Each objective is
audited against the same capacity-matched shuffle
(Table~\ref{tab:reference-free-controls}).

Both substitutions preserve the assignment and suffix verdict while the
matched prefix and middle controls fail.  Thus the retraining reference is not
the sole carrier of the observed structure.  This retrospective evaluation of
complete stored replays does not provide a prospective reference-free
identification procedure for a new queue.

\begin{table}[H]
\centering
\small
\setlength{\tabcolsep}{4pt}
\caption{\textbf{Separate replays carry the same structural verdict under two
objectives that need no retraining reference.}  Entries are medians with
observed $[\min,\max]$ over three complete $k=4$ NegGrad+ surfaces.
$D_{P,2}$ and $D_{M,2}$ are equal-capacity positional controls.
``Separated'' counts Holm-adjusted lower-tail comparisons with $p<0.05$ over
the $1{,}000$ capacity-matched shuffles.}
\label{tab:reference-free-controls}
\begin{tabular}{@{}llrrrrr@{}}
\toprule
Objective & Needs reference & $D_{\mathcal A}$ & $D_{S,2}$ & $D_{P,2}$ & $D_{M,2}$ & Separated \\
\midrule
$J_{\rm ref}$ & Yes & \shortstack[r]{$0.302$\\$[0.265,0.441]$} & \shortstack[r]{$0.221$\\$[0.204,0.286]$} & \shortstack[r]{$2.246$\\$[1.423,2.437]$} & \shortstack[r]{$2.580$\\$[1.550,3.029]$} & 3/3 \\
$J_{\rm NM}$ & No & \shortstack[r]{$0.234$\\$[0.116,0.376]$} & \shortstack[r]{$0.320$\\$[0.189,0.377]$} & \shortstack[r]{$1.634$\\$[1.544,1.794]$} & \shortstack[r]{$1.872$\\$[1.494,2.012]$} & 3/3 \\
$J_{\rm pol}$ & No & \shortstack[r]{$0.190$\\$[0.119,0.773]$} & \shortstack[r]{$0.242$\\$[0.101,0.835]$} & \shortstack[r]{$1.963$\\$[1.693,2.119]$} & \shortstack[r]{$2.157$\\$[1.790,2.428]$} & 3/3 \\
\bottomrule
\end{tabular}%

\end{table}

\subsection{Sequential ordering versus pooled execution}
\label{app:pooling-scope}

This is a policy-scope control, not an ordering result.  The paper fixes a
sequential policy and asks which order to run it in.  A deployment that may
instead aggregate the requests is choosing a different policy.  On the same ten
complete $k=4$ NegGrad+ instances, we executed a single pooled update over the
union of the four request sets under a matched total budget of $48$ steps, with
the same base model, reference and hyperparameters.  All ten pooled executions
satisfy the validity screens, and each reaches a terminal objective below the
best of its own $24$ sequential orders, a median reduction of $37.8\%$
($30.6$--$44.2\%$) against the best sequential order.

When aggregation is available, changing the execution policy can therefore
dominate reordering within a fixed sequential policy.  This does not bear on
the structural claims, which are about the permutation problem a fixed
request-wise policy induces.  That fixed policy is a modeling scope, not a
claim that pooling is unavailable in every deployment.

\subsection{Queue-length diagnostics}
\label{app:queue-diagnostics}

\paragraph{Attenuation across queue lengths.}
\label{app:rms-diagnostic}
Corollary~\ref{cor:queue-uniform} says the quantities to watch as a queue grows
are the contraction and the decision scale, not $k$.  Holding the NegGrad+
policy and its budget fixed, the measured two-request attenuation is nearly
unchanged across $k=4$, $5$ and $8$, with medians $0.069$, $0.071$ and
$0.065$ and between-queue variation smaller than the within-queue range.  This
finite-execution attenuation diagnostic therefore does not drift with $k$ over
the range we execute.

\paragraph{Decision-scale diagnostics.}
\label{app:decision-scale}
Corollary~\ref{cor:queue-uniform} puts every queue-length dependence into
$\mu_1(k)$ of Equation~\ref{eq:decision-scale}, which the complete surfaces
measure directly.  On the full-network settings the last position alone carries
most of the available signal, $\mu_1/\Delta_J=0.89$ at $k=4$ and $0.93$ at
$k=5$, and the scale itself does not shrink with the queue.  $\mu_1$ is $8.31\%$
of the mean objective at $k=4$ and $8.98\%$ at $k=5$.  The decision scale is
therefore non-degenerate over the queue lengths we execute, which is weaker
than the uniform condition $\inf_k\mu_1(k)>0$ that
Corollary~\ref{cor:queue-uniform} assumes.

\section{Limitations and Scope}
\label{app:limitations}

Our analysis conditions on a fixed request-wise policy and terminal objective.
Changing the policy, including pooling requests, is a separate decision and can
dominate reordering in the tested setting
(Appendix~\ref{app:pooling-scope}).  Reference distance is used as a dense
structural objective rather than a complete forgetting or privacy criterion
(Appendices~\ref{app:behavioral-anchoring} and~\ref{app:reference-free}).
Exact optimal orders are more sensitive to execution randomness than the
structural families studied here (Appendix~\ref{app:randomness-robustness}).
Complete permutation surfaces are limited to modest queue sizes, while
larger-$k$ results rely on held-out evaluation
(Appendix~\ref{app:identification}).  Adaptive arrivals, joint policy selection,
and exact complexity under contraction bounded away from one remain open; the
last boundary is detailed in Appendix~\ref{app:assignment-hardness}.

\end{document}